\documentclass[12pt]{article}
\usepackage{amsthm}
\usepackage{amsmath}
\usepackage{natbib}
\usepackage[colorlinks,citecolor=blue,urlcolor=blue,filecolor=blue,backref=page]{hyperref}
\usepackage{graphicx}
\usepackage{multicol}
\usepackage{amssymb}
\usepackage{rotating}
\usepackage{array}
\usepackage[margin=1in]{geometry}

\newcommand{\blind}{1}

\newcommand{\norm}[1]{\left\lVert#1\right\rVert}
\newcommand{\real}[1]{\mathbb{R}}

\theoremstyle{definition}
\newtheorem{definition}{Definition}[section]

\graphicspath{{figs/}}

\begin{document}

\def\spacingset#1{\renewcommand{\baselinestretch}%
{#1}\small\normalsize} \spacingset{1}


\if1\blind
{
  \title{\bf Automatic Detection and Uncertainty Quantification of Landmarks on Elastic Curves}
  \author{Justin Strait, Oksana Chkrebtii, and Sebastian Kurtek\\
    Department of Statistics, The Ohio State University}
  \maketitle
} \fi

\if0\blind
{
  \bigskip
  \bigskip
  \bigskip
  \begin{center}
    {\LARGE\bf Automatic Detection and Uncertainty Quantification of Landmarks on Elastic Curves}
\end{center}
  \medskip
} \fi

\bigskip
\begin{abstract}
A population quantity of interest in statistical shape analysis is the location of landmarks, which are points that aid in reconstructing and representing shapes of objects.  We provide an automated, model-based approach to inferring landmarks given a sample of shape data. The model is formulated based on a linear reconstruction of the shape, passing through the specified points, and a Bayesian inferential approach is described for estimating unknown landmark locations.  The question of how many landmarks to select is addressed in two different ways: (1) by defining a criterion-based approach, and (2) joint estimation of the number of landmarks along with their locations.  Efficient methods for posterior sampling are also discussed. We motivate our approach using several simulated examples, as well as data obtained from applications in computer vision and biology; additionally, we explore placements and associated uncertainty in landmarks for various substructures extracted from magnetic resonance image slices.
\end{abstract}

\noindent%
\vfill

\newpage
\spacingset{1.45} 

\section{Introduction}  \label{Intro}

Shape analysis is an emerging field within statistics due to the necessity of using statistical procedures on shapes of various types of objects. Shape is an important physical property of objects and emerges in many application areas including medical imaging, pattern recognition, computer vision, biometrics, biology, bioinformatics and many others. Statistical procedures on shape spaces are similar in some ways to standard statistical methods developed for univariate and multivariate numerical data. However, developing statistical shape analysis methods requires extra care for several reasons. First, there is no consensus on the choice of shape representation, which determines how these objects are treated mathematically. Many representations have been developed, all of which have their advantages and disadvantages depending on the setting of interest; two of the most common classes are landmark-based and function-based, and are described later in this section. Second, most shape representation spaces are quotients of nonlinear manifolds (which means that performing operations like adding shapes is not straightforward as in a linear space). This is due to the most common definition of shape as an inherited property of an object, which remains unchanged under some transformations (most commonly rotation, scaling and translation). Quotient spaces are required to deem two shapes equivalent when they are only different by this set of transformations. Third, in the case of functional representations of shape, the underlying shape spaces are infinite-dimensional. Thus, any statistical analysis on these spaces requires tools from functional data analysis. In general, statistical shape analysis refers to a set of tools, which can be used for alignment, comparison, averaging, summarization of variability, statistical inference, and other tasks performed on shape spaces.

Initially, the statistical shape analysis community represented the shape of an object using a finite point set comprised of so-called \textit{landmarks}. These ideas were first introduced by \cite{Kendall}, who defined shape as a property of an object which remains unchanged under rigid motion and scaling. The landmark points represent important mathematical (such as curvature) or salient anatomical features of the objects, and are in correspondence across a population of shapes --- this means that, for instance, if a landmark is placed at the tip of a human's nose, then this particular point of the object's outline should be matched with the nose of another human that may be compared to it. In this framework, the entire object is represented by a low-dimensional landmark configuration matrix which is based on the coordinates of the landmarks. After some adjustments to account for the desired shape invariances, one can perform standard multivariate analyses on these shape representation spaces (details provided in \cite{DrydenBook,small_boook:96,dryden-mardia:biometrika:92,bookstein1986}). If landmarks can be located on objects of interest, then this approach provides a low-dimensional representation of shapes for which many statistical tools are readily available. 

As computing technology improved, researchers developed infinite-dimensional, functional representations of shape based on parameterized curves of the objects' outlines.  These representations allow one to model the full structure of the object of interest, but also lead to some additional challenges. Most notably, statistical shape analysis of parameterized curves must also provide invariance to re-parameterization of the curves (in addition to rigid motion and scaling). In other words, any statistical analyses should be exactly the same regardless of the rate at which the curve is traversed. In order to overcome this challenge, \textit{elastic statistical shape analysis} \citep{younes-elastic-distance,SrivESA} was introduced as a parameterization invariant way to compare and model curves. This is accomplished by matching corresponding geometric features across shapes (for instance, the tails of animals being compared), which provides improved results over arc-length parameterization methods \citep{Zahn,klassen-srivastava-etal:04}. \cite{SrivESA} introduced a novel representation of shape called the square-root velocity function (SRVF), which greatly simplifies statistical analysis under the elastic shape analysis paradigm. 

More recently, \cite{LCESA} extended this work to allow hard landmark constraints in the SRVF representation (known as landmark-constrained elastic shape analysis). This new development provides tools for statistical modeling of the full parameterized curve representation of an object's boundary while at the same time respecting given landmark constraints, i.e., by enforcing exact landmark matching.  These methods are useful for comparison of shapes where the entire object is treated as a function, but special points are ``forced" to match (for instance, if a particular shape feature is visible on one object and known but not visible on the other object). Other examples of landmark-constrained elastic shape analysis include the works of \cite{bauer} and \cite{Liu1} where landmarks are treated as soft constraints, i.e., landmarks are used to compute optimal deformations and distances between shapes, but are not necessarily matched exactly.

This work seeks to answer two pertinent questions related to landmark-based shape analysis methods, including landmark-constrained elastic shape analysis. First, in general settings, it is not clear how many landmark points should be selected to represent the shape of interest. Too few landmarks may result in the absence of important features of objects, effectively leading to biased estimation; too many may result in overfitting, a classical statistical problem. Once the number of landmarks is decided, one may also wonder where these landmarks should be located. In the case of anatomical landmarks, the points are usually selected by an expert in the application field, e.g., in medical imaging, doctors manually annotate important anatomical features in an image. However, such an approach is time consuming, expensive and prone to human error. Thus, we propose a novel automatic, model-based approach for answering these two questions under a joint framework. The Bayesian paradigm is a natural approach to infer fixed but unknown landmark locations while accounting for their associated uncertainty.

\subsection{Previous Work}

The proposed automatic landmark detection framework is applicable to any open or closed curve, regardless of type, dimensionality or shape. Automatic landmark detection has been discussed mainly in the presence of specific classes of shapes (e.g., specific anatomy), and primarily from an image analysis perspective. These include the work by \cite{Chen} who focus on the inference of landmarks on X-ray images based on a voting scheme through displaced image patches. Facial landmarks have been the subject of many manuscripts, including those by \cite{Tie}, \cite{Segundo} and \cite{Gilani}.  The latter two focus on curvature-based methods. A more general approach (applicable to an arbitrary class of shapes) was proposed by \cite{Rueda,Rueda2}. \cite{PolyApprox} attempt to characterize the number of sampling points (i.e., landmarks) necessary to approximate shape configurations by polygons; while this method bears some resemblance to what is presented in this manuscript, it is only discussed in the finite-dimensional setting.  In addition, the locations of the landmarks are automatically selected to be equally spaced with respect to arc-length or absolute curvature. Existing frameworks lack a formal underlying statistical model for the sample of shapes, and are based primarily on feature detection and optimization of ad-hoc criteria. To the best of our knowledge, the only mention of a model-based automated landmark detection method in the Bayesian setting was presented by \cite{Domijan}. However, their model was based on image analysis techniques through dependence on pixel values and a segmentation of the image. In contrast, we are interested in finding landmarks directly on the given shapes.

Note that the problem of interest is unrelated to landmark registration of functions found in functional data analysis (see \cite{FDA,CurveStr}), which has the goal of aligning functions based on automatically detected or user-specified landmarks (i.e., local extrema). Our goal is not to register shapes, but to automatically identify these landmarks, which is often quite difficult because identifying local extrema is not trivial for shapes.

The rest of this paper is organized as follows. In order to motivate our framework, we first outline useful background information in Section \ref{MB}. Next, we present our reconstruction-based model for a sample of shapes under the assumption that the number of landmarks is known and fixed (Section \ref{Fix}). In Section \ref{Random}, the original model is extended to include an unknown number of landmarks, which can then be inferred. These models introduce challenging posterior features, which include multimodality and variable dimension. Thus, we also describe specialized methods for sampling from the resulting posterior distributions over the number of landmarks and their locations. Our simulation studies are included in Section \ref{SimE}. Finally, we present several applications in Section \ref{Apps}, and close the paper with a brief summary and a description of future work in Section \ref{Conc}. We provide further discussion on posterior sampling, assessments of MCMC convergence, and additional examples in the Supplementary Materials.

\section{Elastic Shape Analysis Background}	\label{MB}

We present a brief overview of topics in elastic statistical shape analysis relevant to the proposed approach. For further details, please consult \cite{SrivESA}, \cite{KurtekJASA}, and \cite{FSDA}. Let $\beta: \mathcal{D} \rightarrow \mathbb{R}^d$ be an absolutely continuous curve (corresponding to the outline of the object of interest) defined on a domain $\mathcal{D}$. For open curves, the domain is the interval $\mathcal{D}=[0,1]$ and the endpoints, $\beta(0),\ \beta(1)$, do not necessarily agree. Closed curves, defined on the unit circle $\mathcal{D} = \mathbb{S}^1$, will be equivalently represented for ease of exposition by rescaling the domain to $\mathcal{D}=[0,1]$ and enforcing the end-point constraint $\beta(0)=\beta(1)$. In subsequent sections, we focus on the case of planar curves, i.e., $d=2$, but the developed models are readily extended to higher-dimensional curves.  Thus, $\beta$ can be written as $\beta(t)=(\beta_x(t),\beta_y(t))^\top$, where $\beta_x,\ \beta_y$ are coordinate functions mapping $\mathcal{D} \rightarrow \mathbb{R}$.  

In this paper, we consider elastic shapes. Shape is a property of an object which is invariant under rotation, translation, scale, and (in the elastic case) re-parameterization. In other words, the shape of an object which is transformed by the above operations is preserved. A re-parameterization is defined as an element $\gamma$ belonging to the re-parameterization group $\Gamma=\{\gamma:[0,1]\to[0,1]\ |\ \gamma(0)=0,\gamma(1)=1,0<\dot{\gamma}<\infty\}$, where $\dot{\gamma}$ is the time-derivative $\frac{d\gamma}{dt}$. The element $\gamma$ acts on $\beta$ through composition, and effectively changes the rate of traversal of the curve, i.e., $t\mapsto \gamma(t)$. Inference on elastic shapes requires an appropriate metric on the space of elastic shapes, which respects these invariances. The simplest metric for comparing shapes of two curves $\beta_1$ and $\beta_2$ is the $\mathbb{L}^2$ metric defined as $\norm{\beta_1-\beta_2}=\sqrt{\int_{\mathcal{D}} | \beta_1(t)-\beta_2(t)|^2 dt}$, where $| \cdot |$ is the Euclidean norm. However, invariance to re-parameterization is not possible under this metric because the action of the re-parameterization group $\Gamma$ is not distance-preserving, i.e., $\norm{\beta_1-\beta_2} \neq \norm{\beta_1 \circ \gamma - \beta_2 \circ \gamma}$.  To remedy this situation, \cite{SrivESA} consider instead a notion of distance between square-root velocity functions (SRVFs) (see Definition \ref{Def1}).

\theoremstyle{definition}
\begin{definition}   \label{Def1}
The \textit{square-root velocity function} of an absolutely continuous curve $\beta:\mathcal{D} \rightarrow \mathbb{R}^d$  is defined as \[q(t)= \begin{cases} \frac{\dot{\beta}(t)}{\sqrt{\vert \dot{\beta}(t) \vert}} & \text{if} \ \beta \ \text{is differentiable at} \ t \ \text{and} \ \vert \dot{\beta}(t) \vert \neq 0 \\ 0 & \text{otherwise}  \end{cases} , \] where $\dot{\beta}$ denotes the time-derivative of $\beta$ (taken over each coordinate function separately) and $\vert \cdot \vert$ is the standard Euclidean norm in $\mathbb{R}^d$.
\end{definition}

There are several benefits to representing a curve by its SRVF.  First of all, $q$ encodes the direction $\Big( \frac{\dot{\beta}(t)}{\vert \dot{\beta}(t) \vert} = \frac{q(t)}{\vert q(t) \vert} \Big)$ and instantaneous speed ($\vert \dot{\beta}(t) \vert = \vert q(t) \vert^2$) of $\beta$.  In fact, given the assumption of absolute continuity and a starting point $\beta(0)$, there is a smooth bijective mapping between $q$ and $\beta$ (\cite{DynPro}); a mapping from SRVF $q$ to its $\beta$ is given by $\beta(t)=\beta(0)+\int_0^t q(s) \vert q(s) \vert ds$.  Notice that the SRVF is automatically invariant to translations due to the sole dependence on $\dot{\beta}$.  This representation is also useful because it is valid for both open and closed curves of any dimension $d$.  Most importantly to the context of shape analysis, the $\mathbb{L}^2$ distance between SRVFs is equivalent to the elastic distance between the original curves, which measures the amount of bending and stretching required to deform one shape into the other (see \cite{younes-elastic-distance,joshi-klassen-cvpr:07,SrivESA,KurtekJASA} for more details). The elastic metric, and thus the $\mathbb{L}^2$ distance between SRVFs, is invariant to the re-parameterization of curves. One can additionally impose the scale and rotation invariances by restricting the space of SRVFs to a Hilbert sphere and considering quotient structures \citep{KurtekJASA}. In this paper, we consider the $\mathbb{L}^2$ distance between SRVFs as the notion of distance between shapes.

\section{Model for Detection of a Fixed Number of Landmarks}	\label{Fix}

This section assumes that the researcher knows how many landmarks are to be selected for a population of shapes. A discussion of all parts of the model, as well as methodology for sampling from the posterior distribution over the landmark locations are presented. In Section \ref{Random}, we extend the model to the case where the appropriate number of landmarks is unknown.

\subsection{Model Specification for Shape Data}		\label{Bayes}

Following existing landmark-based statistical shape analysis methods, we first consider the problem of identifying a fixed number of landmarks $k$, denoted by their domain locations $\boldsymbol{\theta}=(\theta_1,\ldots,\theta_k) \in \mathcal{D}^k$ for a population of shapes, subject to the constraint $\theta_1<\ldots<\theta_k$. We assume that this population is homogeneous, meaning that an arc-length parameterization of shape is sufficient and shape registration is not necessary. Let $\beta_1,\ldots,\beta_M: \mathcal{D} \rightarrow \mathbb{R}^2$ be a sample of curves formed from the outlines of $M$ objects from the population. As a pre-processing step, we re-scale the original curves to unit length in order to give equal weight to each shape in the sample; this is necessary due to the potential for one shape in a sample to dominate inference if the size of the object is quite large.

To specify the likelihood of data $\boldsymbol{\beta}$ given a set of landmarks $\boldsymbol{\theta}$, we first consider reconstruction of shapes using landmark locations. We choose reconstructions of the $m^{th}$ curve in the sample, $\beta_m$, via a linear interpolation constructed by piecewise segments $L_m(t;\theta_i,\theta_{i+1})$ passing through the landmark locations. The expression for the linear interpolator segment between $\theta_i$ and $\theta_{i+1}$ for $i=1,\ldots,k-1$ is given by: 

\begin{equation} \label{eq:LinInt}
L_m(t;\theta_i,\theta_{i+1})=\bigg(1-\frac{t-\theta_i}{\theta_{i+1}-\theta_i}\bigg)\beta_m(\theta_i)+\bigg(\frac{t-\theta_i}{\theta_{i+1}-\theta_i}\bigg)\beta_m(\theta_{i+1}), \quad \theta_i \leq t < \theta_{i+1}.
\end{equation}

\noindent Note that Equation \ref{eq:LinInt} only gives an expression for the segment between landmarks; this does not define the entire interpolator.  For open curves, the linear reconstructions must additionally connect to $\beta_m(0)$ and $\beta_m(1)$. Thus, two additional segments are required: (1) one connecting $\beta_m(0)$ and $\beta_m(\theta_1)$, and (2) one connecting $\beta_m(\theta_k)$ and $\beta_m(1)$.  The expression for these segments is identical to Equation \ref{eq:LinInt}, treating the starting point of these segments as $\theta_i$ and ending point of these segments as $\theta_{i+1}$.  For closed curves, we know $\beta_m(0)=\beta_m(1)$, so the linear reconstruction must also be closed.  This is guaranteed by forming a segment connecting $\beta_m(\theta_k)$ and $\beta_m(\theta_1)$, again using Equation \ref{eq:LinInt}.  For the remainder of the paper, we suppress the $t$ input and write $L_m(\boldsymbol{\theta})$ to represent the full linear interpolator for the $m^{th}$ curve, constructed by joining these piecewise linear interpolator segments. 

We observe a random sample of $M$ shapes $\boldsymbol{\beta}=(\beta_1,\ldots,\beta_M)^\top$, with corresponding SRVFs $q_{\beta_1},\ldots,q_{\beta_M}$.  Given landmark locations $\boldsymbol{\theta}$, we obtain the SRVFs of their linear reconstructions $q_{L_1(\boldsymbol{\theta})},\ldots,q_{L_M(\boldsymbol{\theta})}$. Using these ideas, we can model the SRVFs $q_{\beta_1}, \ldots, q_{\beta_M}$ of our shape data as follows:
\begin{equation}
q_{\beta_m} = q_{L_m(\boldsymbol{\theta})} + \varepsilon_m, \quad m = 1,\ldots,M ,
\end{equation}
where $\varepsilon$ is a Gaussian process on $\mathcal{D}$ with zero mean function and covariance function $C$. Thus, we can write, 
\begin{equation} \label{eq:GP}
q_{\beta_m}   \mid  q_{L_m(\boldsymbol{\theta})}, \boldsymbol{\theta} \sim  \mathcal{GP} \left( q_{L_m(\boldsymbol{\theta})} , C\right).
\end{equation}
Here, we choose the covariance function $C(t_1,t_2) = \delta_{t_1,t_2}(2 \kappa)^{-1}$, where $\delta$ is the Dirac delta function that equals one if $t_1 = t_2$ and is zero otherwise. This can be readily generalized to encode, for example, correlation between components of $\varepsilon_m$, or its desired smoothness.

\subsubsection{Likelihood} \label{Lik}

\begin{figure}[!t]
\begin{center}
\begin{tabular}{|c|c|}
\hline
\includegraphics[width=0.3\textwidth]{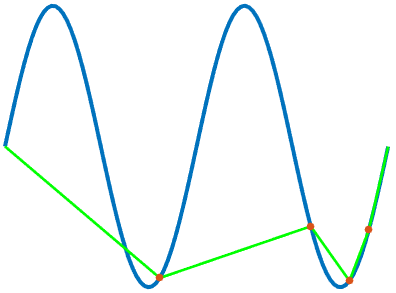}&\includegraphics[width=0.3\textwidth]{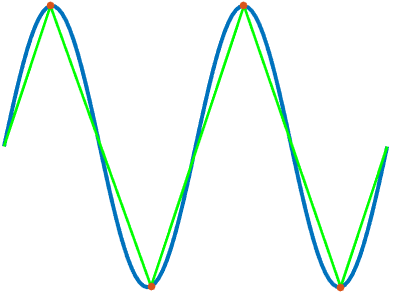}\\
\hline
$d^2(\beta^{(N)}_m,L^{(N)}_m( \boldsymbol{\theta}))=0.8429$ & $d^2(\beta^{(N)}_m,L^{(N)}_m( \boldsymbol{\theta}))=0.0739$\\
\hline
\end{tabular}
\end{center}
\caption{Poor (left) and good (right) reconstructions of $\beta_m$ (defined as a two-dimensional curve in Section \ref{SimE}), with squared reconstruction errors reported below. $\beta_m$ is in blue, $L_m(\boldsymbol{\theta})$ is in green, with landmarks specified by $\boldsymbol{s}$ shown as red dots.}
\label{fig:BadGood}
\end{figure}

For a curve $\beta_m$ with reconstruction $L_m( \boldsymbol{\theta})$ we first compute the corresponding SRVFs $q_{\beta_m}$ and $q_{L_m( \boldsymbol{\theta})}$. Then, assuming that $q_{\beta_m}$ and $q_{L_m( \boldsymbol{\theta})}$ are discretized using $N$ points (call these \textit{evaluation points}), we define the reconstruction error as

\begin{equation}  \label{eq:RecErr}
d(\beta^{(N)}_m,L^{(N)}_m( \boldsymbol{\theta}))=|\text{vec}(q^{(N)}_{\beta_m}-q^{(N)}_{L_m( \boldsymbol{\theta})})| ,
\end{equation}
where $\text{vec}$ is the vectorize operator which forms a vector of size $2N$ by vertically stacking the $x$ and $y$ coordinates of the SRVFs, $|\cdot|$ is the Euclidean norm in $\mathbb{R}^{2N}$, and $f^{(N)}$ denotes the function $f$ discretized using $N$ points.  (In general, if we are dealing with a $d$-dimensional curve, the vectorize operator will form a vector of size $dN$.)  A small value of $d(\beta^{(N)}_m,L^{(N)}_m( \boldsymbol{\theta}))$ indicates that the landmarks $\boldsymbol{\theta}$ yield an accurate reconstruction of the $m^{th}$ curve, and the landmarks approximate the full object well. Figure \ref{fig:BadGood} shows two landmark configurations; the one on the left results in a large reconstruction error $d(\beta^{(N)}_m,L^{(N)}_m( \boldsymbol{\theta}))$, as the chosen landmarks do not provide a faithful reconstruction of the original curve.  The one on the right is much better, and thus has a smaller reconstruction error $d(\beta^{(N)}_m,L^{(N)}_m( \boldsymbol{\theta}))$.

Discretizing Equation \ref{eq:GP} yields,
\begin{equation}
\text{vec}\big(q^{(N)}_{\beta_1}-q^{(N)}_{L_1(\boldsymbol{\theta})}\big),\ldots,\text{vec}\big(q^{(N)}_{\beta_M}-q^{(N)}_{L_M(\boldsymbol{\theta})}\big)|\boldsymbol{\theta},\kappa \overset{iid}{\sim} N\bigg(0_{2N},\frac{1}{2\kappa}I_{2N}\bigg),
\end{equation}
where $\kappa=\frac{1}{2\sigma^2}$ is a precision parameter. The likelihood function for the data $\boldsymbol{\beta}^{(N)}=(\beta_1^{(N)},\ldots,\beta_M^{(N)})^\top$ is then given by,
\begin{equation}
f(\boldsymbol{\beta}^{(N)}|\boldsymbol{\theta},\kappa)=\pi^{-NM} \kappa^{NM} \exp \Bigg(-\kappa \sum_{m=1}^M d^2(\beta^{(N)}_m,L^{(N)}_m(\boldsymbol{\theta})) \Bigg),
\end{equation}
where $d(\beta^{(N)}_m,L^{(N)}_m(\boldsymbol{\theta}))$ is defined in Equation \ref{eq:RecErr}. For a $d$-dimensional curve, the normal model is still appropriate, where $dN$ replaces $2N$ in the mean and variance.

Notice that this likelihood is defined in terms of the \emph{SRVFs of the original curves and their linear reconstructions}, rather than the original coordinates of the curves and their reconstructions.  Since the SRVF is related to the instantaneous velocity of the curve, it is reasonable to use a normal model, as both negative and positive values of $q^{(N)}_{\beta_m}-q^{(N)}_{L_m(\boldsymbol{\theta})}$ are feasible at each of the $N$ evaluation points.  This likelihood model is similar to that of \cite{Cheng} and \cite{KurtekEJS}, where it was used for registration of functional data.

\subsubsection{Prior}	\label{Prior}

Next, we specify prior distributions on $\kappa$ and $\boldsymbol{\theta}$.  A priori, assume $\kappa$ and $\boldsymbol{\theta}$ are independent: $\pi(\kappa,\boldsymbol{\theta})=\pi(\kappa)\pi(\boldsymbol{\theta})$. Because $\kappa$ is a precision parameter (and a nuisance parameter), we specify a prior that is conditionally conjugate under our normal model (see e.g., \cite{GelmanEtAl2004}):
\begin{equation}
\kappa \sim \text{Gamma}(a,b).
\end{equation}

Prior specification for $\boldsymbol{\theta}$ is quite challenging due to the ordering constraint on its components.  In order to simplify this task, we transform $\boldsymbol{\theta}$ to a vector of consecutive differences between landmarks, denoted by $\boldsymbol{s}$. The dimension of $\boldsymbol{s}$ depends on whether we are detecting landmark locations on open or closed curves. For open curves, we define the components $s_i=\theta_{i+1}-\theta_i$ for $i=1,\ldots,k-1$; we set $s_0=\theta_1$ and $s_k=1-\theta_k$ (as the linear reconstruction is required to pass through the start and end points of the curve), and let $\boldsymbol{s}=(s_0,s_1,\ldots,s_k)$ which is $(k+1)$-dimensional. For closed curves, the components are still defined as $s_i=\theta_{i+1}-\theta_i$ for $i=1,\ldots,k-1$; however, we let $s_k=(\theta_1-\theta_k)\ \text{mod} \ 1$. There is a one-to-one correspondence between $\boldsymbol{\theta}$ and $\boldsymbol{s}$ for open curves; for closed curves, if a starting point along the curve is designated, then a one-to-one correspondence is also achieved. Thus, we proceed by using $\boldsymbol{s}$, and then recover $\boldsymbol{\theta}$ for inferential purposes. For notational simplicity, any notation which depends on $\boldsymbol{\theta}$ may also be written to depend on $\boldsymbol{s}$ instead (i.e., $d(\beta^{(N)}_m,L^{(N)}_m(\boldsymbol{\theta}))$ is equivalent to $d(\beta^{(N)}_m,L^{(N)}_m(\boldsymbol{s}))$).

Thus, we place a prior on $\boldsymbol{s}$. Notice that $\sum_{i} s_i = 1$ and $s_i>0$ for all $i$ for both open and closed curves. In addition, by construction, $\boldsymbol{s}$ does not have the ordering constraint on its components. Thus, a natural prior for $\boldsymbol{s}$ is the Dirichlet distribution with concentration parameter $\alpha$:
\begin{equation}
\boldsymbol{s} \sim \text{Dir}(\alpha\boldsymbol{1}),
\end{equation}
where $\boldsymbol{1}$ is a vector of ones with dimension equal to $k+1$ for open curves and $k$ for closed curves. Choice of prior hyperparameters is required. Selecting $a=1, \ b=0.01$ for the prior on $\kappa$ is very weakly informative, and we show in Table \ref{tab:Sensitivity} that inference is robust to reasonable choices of $a$ and $b$. For the prior on $\boldsymbol{s}$, we select $\alpha=1$; this choice does not favor a particular spacing between landmarks (i.e., does not favor landmark clustering), and is uniform across the support of $\boldsymbol{s}$.

\subsubsection{Posterior}	\label{Posterior}

The density of the posterior distribution over landmark spacing $\boldsymbol{s}$ given the data $\boldsymbol{\beta}^{(N)}$ is denoted by $\pi(\boldsymbol{s}|\boldsymbol{\beta}^{(N)})$. The precision parameter $\kappa$ is not of direct interest to us (since our goal is to solely infer landmark locations), so we compute the marginal likelihood,
\begin{equation}
f(\boldsymbol{\beta}^{(N)}|\boldsymbol{s})=\int_{\mathbb{R}_+}f(\boldsymbol{\beta}^{(N)}|\boldsymbol{s},\kappa)\pi(\kappa) d\kappa = \frac{\pi^{-NM} \Gamma(a+NM) b^a}{\Gamma(a)\bigg(b+\sum_{m=1}^M d^2(\beta^{(N)}_m,L^{(N)}_m(\boldsymbol{s})) \bigg)^{a+NM}},
\end{equation}
and use it to obtain the posterior density,
\begin{equation}
\pi(\boldsymbol{s}|\boldsymbol{\beta}^{(N)}) \propto f(\boldsymbol{\beta}^{(N)}|\boldsymbol{s}) \pi(\boldsymbol{s}).
\end{equation}
Note that if $\kappa$ is of interest to the researcher, then the algorithm described in Section \ref{MCMC} can be implemented, with an additional Gibbs step to sample from the full conditional of $\kappa$. Posterior samples of $\boldsymbol{s}$ can be transformed to $\boldsymbol{\theta}$ as described in Section \ref{Prior}. As posterior functionals of interest are not analytically tractable, inference will be based on approximations computed from Markov chain Monte Carlo (MCMC) samples. Section 1 of the Supplementary Materials provides an alternative approach based on importance sampling.  While MCMC is more appropriate for this model due to poor scaling of importance sampling with respect to the number of curves and their sampling density, importance sampling can still be useful for quick posterior estimation of landmark location means and maxima a posteriori (MAP).

\subsection{Sampling via MCMC}		\label{MCMC}
Estimates based on MCMC samples enable Bayesian inference when posterior functionals of interest are not available analytically. In the case of this model, as the number of curves $M$ increases, the posterior exhibits a complex correlation structure and multimodality. For posterior inference on the locations of a fixed number of landmarks, we use the random walk Metropolis algorithm. We initialize the algorithm by sampling $\boldsymbol{s}^{[0]}$ from $\pi(\boldsymbol{s})$. The superscript $^{[t]}$ will denote the state of the Markov chain at iteration $t$. For a given MCMC iteration $t$, a proposal vector of landmarks is generated by selecting the $j^{th}$ component of the landmark vector $\boldsymbol{\theta}^{[t]}$ and applying a symmetric proposal distribution $h$. For our implementation, let $h$ be a normal probability density function, with mean $\theta_j^{[t]}$ and variance $v$. This symmetric proposal mechanism yields a proposed vector $\boldsymbol{s}^*$. The Metropolis acceptance ratio is then defined as:
\begin{equation}
\alpha(\boldsymbol{s}^{[t]},\boldsymbol{s}^*)=\frac{\pi(\boldsymbol{s}^*|\boldsymbol{\beta}^{(N)})}{\pi(\boldsymbol{s}^{[t]}|\boldsymbol{\beta}^{(N)})}=\frac{f(\boldsymbol{\beta}^{(N)}|\boldsymbol{s}^*)\pi(\boldsymbol{s}^*)}{f(\boldsymbol{\beta}^{(N)}|\boldsymbol{s}^{[t]})\pi(\boldsymbol{s}^{[t]})}.
\end{equation}
The proposal $\boldsymbol{s}^*$ is accepted with probability $\min\{1,\alpha(\boldsymbol{s}^{[t]},\boldsymbol{s}^*)\}$. As this procedure only updates one parameter at a time, convergence can require a large number of iterations; this is feasible because the likelihood is easy to evaluate. The algorithm is monitored for convergence, and approximate posterior samples are obtained after a suitable burn-in period with a thinning step to reduce autocorrelations.

MCMC is an efficient approach for joint landmark posterior inference for a large number of curves or a high sampling density of curves, as methods like importance sampling suffer in these settings due to the issues discussed in Section 1 of the Supplementary Materials.  However, care must be taken in selecting proposals to ensure that the chain traverses the parameter space efficiently; this can be especially important in the specified model, as the posterior is often multimodal. In all cases, samples which violate the ordering assumption of $\boldsymbol{\theta}$ are automatically rejected; this is evident in the specification of the prior on $\boldsymbol{s}$. Care must also be taken when dealing with closed curves: since there is no designated start or end point, proposals must be allowed to wrap around the circular curve domain. We discuss the implementation for closed curves in more detail in the next section.

\subsection{Implementation for Closed Curves}	\label{Imp}

\begin{figure}[!t]
\begin{center}
\includegraphics[width=4.5in]{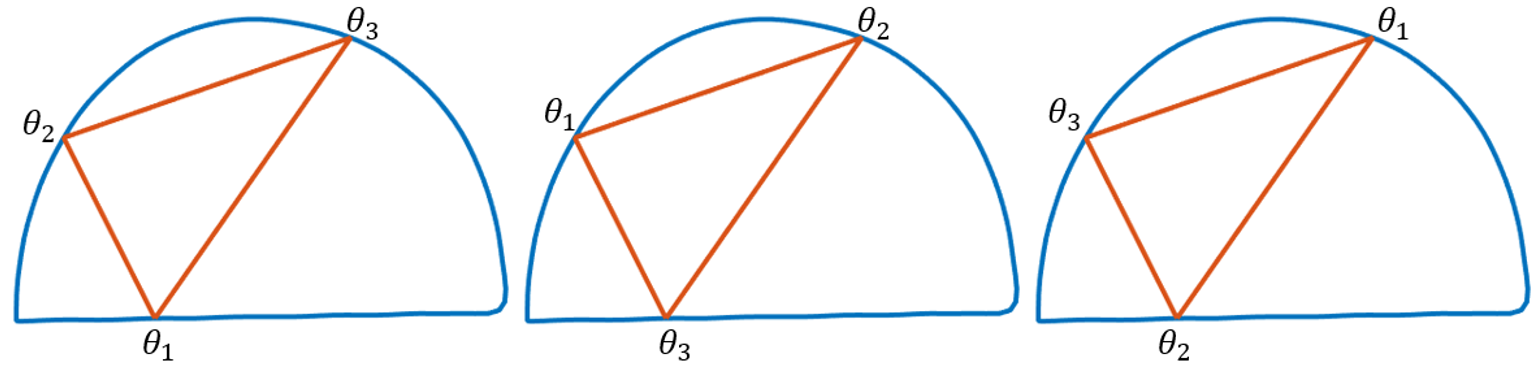}\\
\end{center}
\caption{An example of the identifiability issue encountered when detecting landmarks on closed curves, which is resolved by deeming these three landmark configurations equivalent under our model specification.}
\label{fig:Iden}
\end{figure}

The issue of identifiability arises when dealing with closed curves; the domain is $\mathbb{S}^1$, which has no natural start or end point. Thus, we must find a point in the domain that can be identified with $t=0$. Figure  \ref{fig:Iden} illustrates this with $k=3$ landmarks for a half-circle; the reconstruction is invariant to how the landmarks $\theta_1,\ \theta_2,\ \theta_3$ are labeled. This means that the model is exchangeable with respect to the ordering of $\boldsymbol{\theta}$.

To address this issue, we designate a reference point with parameter value $\theta_0$ to be the point of maximal curvature on the first shape in the sample. Then, we pre-process the entire collection of curves as follows. We shift the order of points for the first curve such that $t=0$ is identified with $\theta_0$. Then, all subsequent curves are aligned to the first curve by finding the ordering of points which minimizes the SRVF distance to the first curve. Note that this is not a registration step; we are systematically defining a starting point on each curve, since this is not well-defined for a given sample of closed curves. In order to visualize the posterior samples for closed curves, we post-process them to lie on the re-scaled unit circle by eliminating the boundary between $\theta=0$ and $\theta=1$ as follows. We align all $n$ posterior samples according to the set of locations for the first posterior sample $\boldsymbol{\theta}_1$. That is, for samples $\boldsymbol{\theta}_i$ for $i=2,\ldots,n$, we compute the distance between corresponding components $j=1,\ldots,k$ using the following circular metric: $$d^{[j]}(\boldsymbol{\theta}_i,\boldsymbol{\theta}_1)=\min\{|\theta^{[j]}_i-\theta^{[j]}_1|,|\theta^{[j]}_i-1-\theta^{[j]}_1|,|\theta^{[j]}_i+1-\theta^{[j]}_1|\}.$$ Then, we find the landmark ordering (using circular permutations) which minimizes $d(\boldsymbol{\theta}_i,\boldsymbol{\theta}_1)=\sum_{j=1}^k d^{[j]}(\boldsymbol{\theta}_i,\boldsymbol{\theta}_1)$. Note that this alignment process is done using the full configuration of $\boldsymbol{\theta}$. We have found this procedure to be robust to the choice of the first posterior sample based on many simulations and real data examples. We note that the most appropriate approach would be to jointly compute the posterior mean of all samples while modding out by $\mathbb{S}^1$. Further investigation of this idea is left as future work.

\section{Model for Joint Estimation of the Number of Landmarks and Their Locations}	\label{Random}

Addressing how many landmarks to select on a given set of curves is a complex task, and is akin to a model selection problem found in many facets of statistics, where ``the number of things you don't know is one of the things you don't know" (\cite{RichardsonGreen1997}). There are generally two ways in which statisticians approach this problem. One way is to develop a criterion which must be optimized while making sure to not ``overfit" the model. The other is to treat the number of parameters as unknown and infer it from the data. In this section, we discuss two such approaches for selecting $k$, the number of landmarks.

\subsection{Distance-Based Criterion}	\label{ASDC}

The criterion-based method for selecting $k$ is borrowed from dimension reduction problems for high-dimensional data. One example of this is principal component analysis (PCA), which forms a much lower-dimensional space of uncorrelated modes of variation; these modes of variation are first ordered by proportion of variability explained.  The number of components is selected by choosing a cut-off where, at a certain point, the percentage of variation begins to ``level off", as adding additional components becomes unnecessary. Typically, the number of components selected is chosen based on the ``elbow" of the plot of percent variation vs. the number of components (known as a scree plot).

We use a similar approach to choose the number of landmarks $k$. For each value of $k$ considered, we draw posterior samples $\boldsymbol{\theta}_1,\ldots,\boldsymbol{\theta}_n$ (notice that $n$ is the number of posterior samples of $\boldsymbol{\theta}$ generated, each of which is a $k$-dimensional vector). For a posterior sample indexed by $i$, we form the linear reconstructions $L^{(N)}_{1i},\ldots,L^{(N)}_{Mi}$ (for the $M$ curves in the data), and compute the average cumulative squared distance $d_k^2=\frac{1}{n}\sum_{i=1}^n\sum_{m=1}^M d^2(\beta^{(N)}_{m},L^{(N)}_{mi}(\boldsymbol{s}))$. We repeat this procedure for many values of $k$ and plot the resulting values; we expect smaller $d_k^2$ for larger values of $k$, as linear reconstructions improve as the number of landmarks increases. Then, $k$ can be chosen at the ``elbow" of this curve, which is the point at which adding additional landmarks does not lead to a substantial reduction in the reconstruction errors.

The distance criterion is intuitive, but is not without issues. The goal of building this model is to automatically select the number of landmarks without any manual selection. Using a plot of $d_k^2$ requires the user to identify the ``elbow", and select that value as the desired number of landmarks. This choice may not always be immediately obvious, and prevents automatic inference of a very useful parameter. It also leaves the user unable to quantify uncertainty in choosing $k$. Thus, it may instead be better to let $k$ be unknown and build it into the Bayesian model.

\subsection{Extension of the Landmark Detection Model to Unknown $k$}	\label{ext}

Unknown $k$ is considered by conditionally specifying an additional level in the Bayesian hierarchical model. The likelihood, now $\boldsymbol{\beta}^{(N)}|\boldsymbol{s},k$, is identical to the likelihood $\boldsymbol{\beta}^{(N)}|\boldsymbol{s}$ in Section \ref{Lik} (after marginalizing over the prior on $\kappa$, which is assumed independent of $k$). The prior on the locations $\boldsymbol{s}$ and their number $k$ is specified as:
\begin{equation}
\pi(\boldsymbol{s},k)=\pi(\boldsymbol{s}|k)\pi(k).
\end{equation}
The prior $\boldsymbol{s}|k$ still follows a Dirichlet distribution (as described in Section \ref{Prior}), where the dimension of the concentration parameter vector depends on $k$. We must specify a prior on $k$ as well. Note that for open curves, $k \geq 1$ (in order to get a valid reconstruction); however, for closed curves, there is no start or end point, so $k \geq 3$. To account for this, we choose as the prior for $k$ a shifted Poisson distribution: we assume $k=\nu+1$ and $k=\nu+3$ for open and closed curves, respectively. Then, the prior on $\nu$ is given by:
\begin{equation}
\nu \sim \text{Poisson}(\lambda).
\end{equation}
The shift guarantees that prior probabilities are greater than zero for the appropriate values of $k$ only. Selection of $\lambda$ is an interesting and difficult problem; the goal for the presented model is to select a $k$-dimensional set of landmark locations, where $k$ is relatively small. Thus, we treat $\lambda$ as a regularization parameter.  Because the likelihood only depends on reconstruction error, adding more landmarks will generally increase the likelihood relative to the prior. Thus, to avoid overfitting, the prior on $k$ can be chosen to place most of its mass very close to zero to penalize choosing high values of $k$. Varying $\lambda$ will therefore yield a path of posterior inference solutions; the dependence of posterior inference on the choice of $\lambda$ is shown in Section $\ref{CompInf}$.

\subsection{Posterior Sampling Using Reversible Jump MCMC}	\label{RJMCMC}
Treating $k$ as unknown complicates posterior inference on $\boldsymbol{s},k|\boldsymbol{\beta}^{(N)}$ due to the dependence of the dimensionality of $\boldsymbol{s}$ on $k$: different values of $k$ result in a different number of parameters to infer. Standard MCMC methods are defined on parameter spaces of fixed dimension. Dependent proposals between parameter spaces of different dimension $k$ can be made via reversible jump MCMC (RJMCMC) \citep{Green1995,RichardsonGreen1997,10.2307/2346184}. This type of procedure is commonly used in model selection problems, where one wants to infer model parameters as well as the number of parameters. In particular, the birth-death form of RJMCMC proposes a new parameter vector by first randomly choosing to increase the dimension of the parameter space by one (a birth), decrease the dimension by one (a death), or keep the dimension the same (a stay). In the case of a birth, a new component is added to the model according to a chosen distribution. Similarly, a component is ``killed" through random selection. This extra step of selecting a move type and developing the proposal based on the selected move is built into the Metropolis-Hastings ratio. Section 2 of the Supplementary Materials describes the RJMCMC procedure for the proposed automatic landmark detection model, as well as further details about specific steps within the algorithm.

\section{Simulation Experiments}	\label{SimE}

\subsection{Selection of a Fixed Number of Landmarks}	\label{OSC}
In order to test various properties of the proposed model, we construct a simple shape based on a sine curve with well-defined peaks and valleys. Consider the curve $\beta(t) = [t,\ \sin(4\pi t)]^\top,\ 0 \leq t \leq 1$, which features two peaks and two valleys, each of which appear to be optimal locations for landmark placement due to low reconstruction error (yielding a total of $k=4$ landmarks). We first begin by drawing posterior samples under the fixed $k=4$ model (with $N=200$ evaluation points). We use the random walk Metropolis algorithm, as described in Section \ref{MCMC}.  We specify $a=1,\ b=0.01$ in the prior for $\kappa$, $\alpha=1$ in the prior for $\boldsymbol{s}$, and a variance of $v=0.02$ for the normal proposal density. The chain is run for $10^6$ iterations; the first ten percent is discarded as burn-in, and the remaining sample is thinned by 100 to reduce autocorrelations.  Trace plots used to diagnose convergence for this example are shown in Section 3 of the Supplementary Materials.

\begin{figure}[!t]
\begin{center}
    \begin{tabular}{|c|c|}
    \hline
    \includegraphics[width=2in]{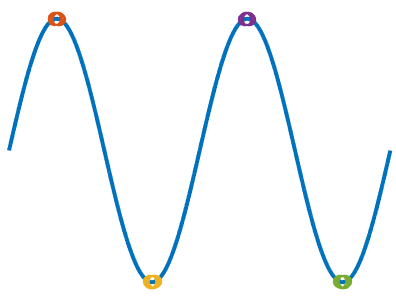}&\includegraphics[width=1.7in]{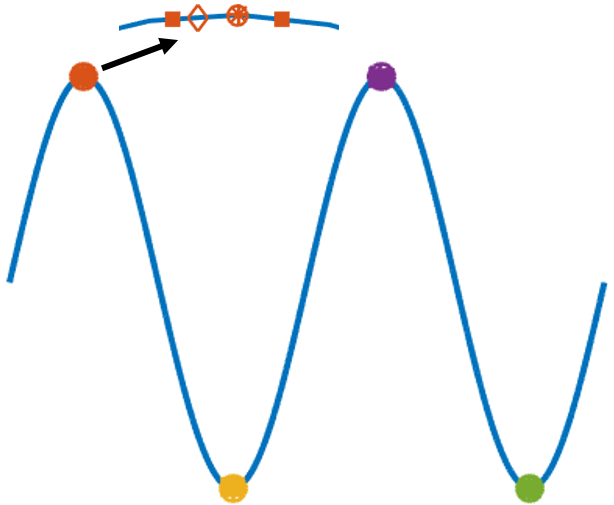}\\
    \hline
    \end{tabular}
    \begin{tabular}{|c|c|c|c|}
    \hline
    \includegraphics[width=1.1in]{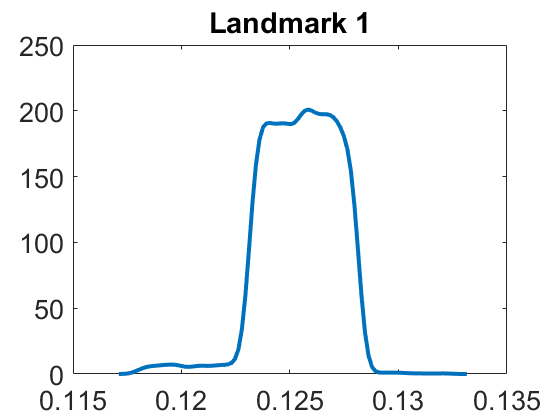} & \includegraphics[width=1.1in]{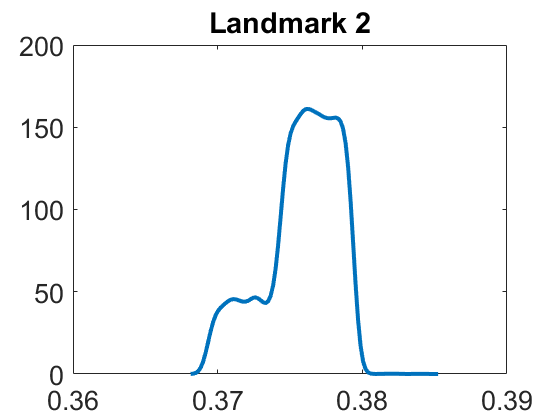}&
    \includegraphics[width=1.1in]{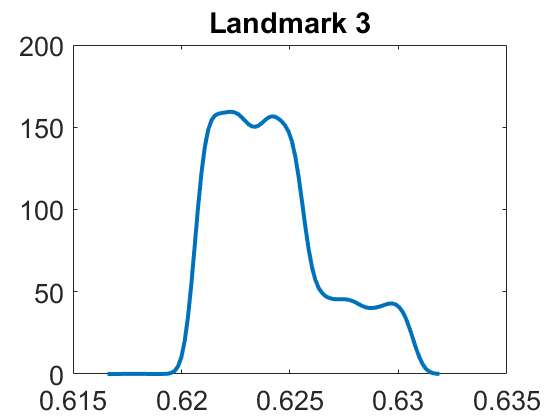} & \includegraphics[width=1.1in]{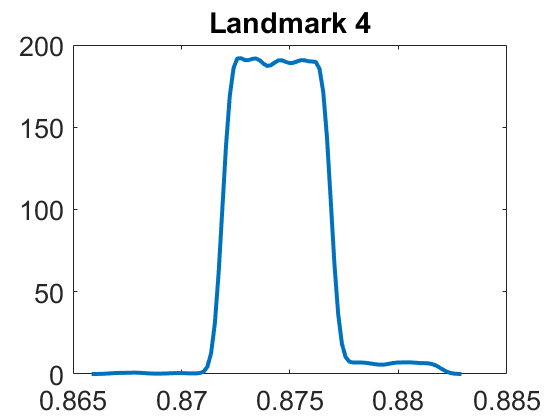}\\
    \hline
	\end{tabular}
\begin{tabular}{|c|c|}
\hline
\textbf{Mean:} $\overline{\boldsymbol{\theta}}=(0.1255,0.3758,0.6242,0.8745)$\\
\hline
\textbf{Median:} $\textrm{med}(\boldsymbol{\theta})=(0.1256,0.3762,0.6238,0.8745)$\\
\hline
\textbf{MAP:} $\boldsymbol{\theta}_{MAP}=(0.1233,0.3748,0.6207,0.8721)$\\
\hline
\end{tabular}
\end{center}
\caption{Top left: Curve $\beta$ with posterior landmark locations obtained using MCMC: red = 1, yellow = 2, purple = 3, green = 4. Top right: Curve $\beta$ with posterior landmark summaries: circle = mean, asterisk = median, diamond = MAP, squares = 95\% credible interval. Bottom: Density plots of marginal posterior samples $\theta_i|\beta^{(N)},\ i=1,\ldots,4$. The table shows posterior summaries of $\boldsymbol{\theta}$.}
\label{fig:ToyExLand}
\end{figure}

The top left panel of Figure \ref{fig:ToyExLand} shows the original curve with samples from the marginal posteriors of landmark locations plotted on top.  The posterior samples from $\boldsymbol{\theta}|\beta^{(N)}$ obtained using our model coincide with the peaks and valleys of $\beta$ as expected. Posterior uncertainty for each landmark is illustrated in the density plots in the bottom of Figure \ref{fig:ToyExLand}. Each density is fairly concentrated, indicating high confidence in identifying the four landmark locations.  Standard posterior summaries can also be computed for $\boldsymbol{\theta}$. The top right panel of Figure \ref{fig:ToyExLand} shows the mean, median, MAP, and 95\% credible intervals for each component of $\boldsymbol{\theta}$. The mean and median are very similar; the MAP estimate is a little bit different, due to the complex dependencies in the landmark locations.  The 95\% credible intervals are narrow and disjoint, indicating precisely estimated landmark locations.

\subsection{Model Invariance to Shape-preserving Transformations}	\label{Trans}
\begin{figure}[!t]
\begin{center}
    \begin{tabular}{|c|c|c|c|c|c|}
    \hline
        \begin{sideways}Re-scaled \end{sideways}&\includegraphics[width=0.8in]{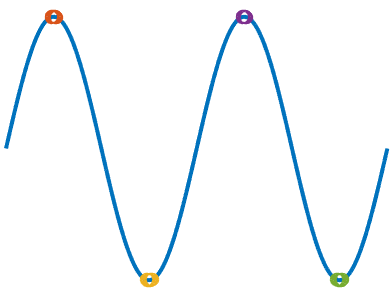}&\includegraphics[width=0.82in]{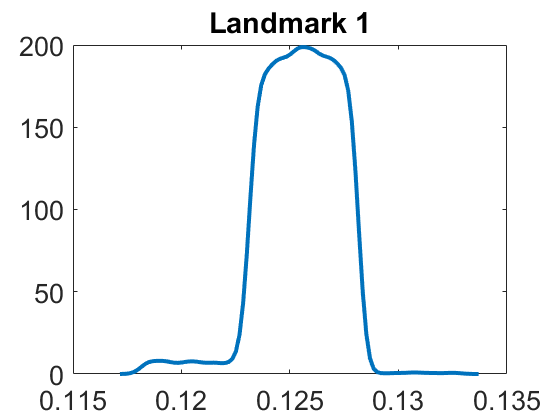}&\includegraphics[width=0.82in]{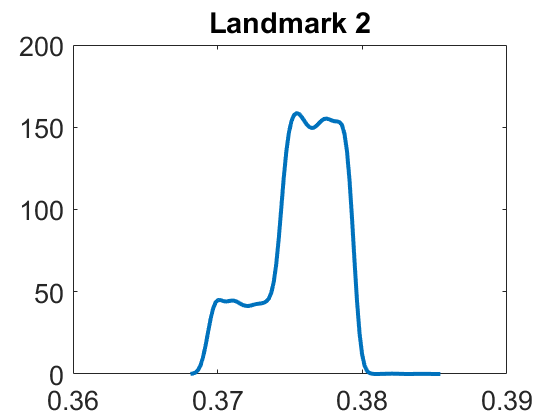}&\includegraphics[width=0.82in]{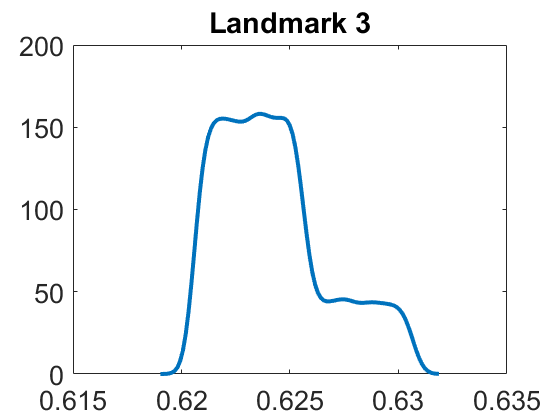}&\includegraphics[width=0.82in]{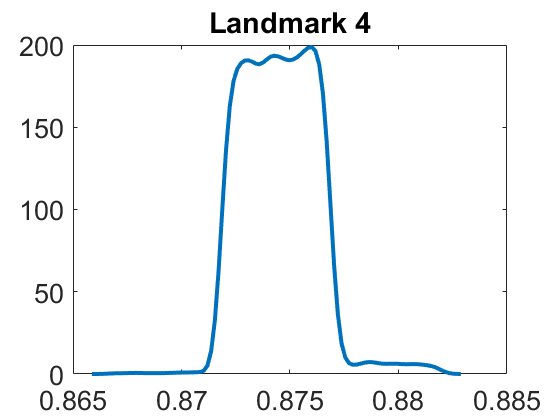}\\
    \hline
    \begin{sideways} Rotated \end{sideways}&\includegraphics[width=0.8in]{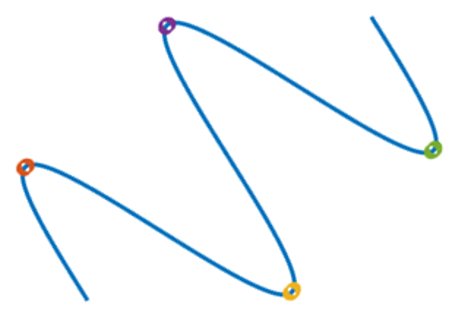}&\includegraphics[width=0.82in]{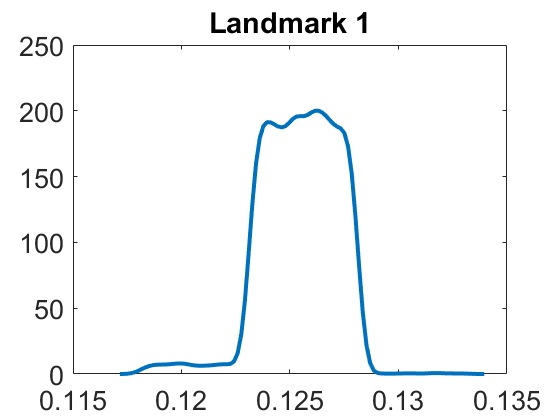}&\includegraphics[width=0.82in]{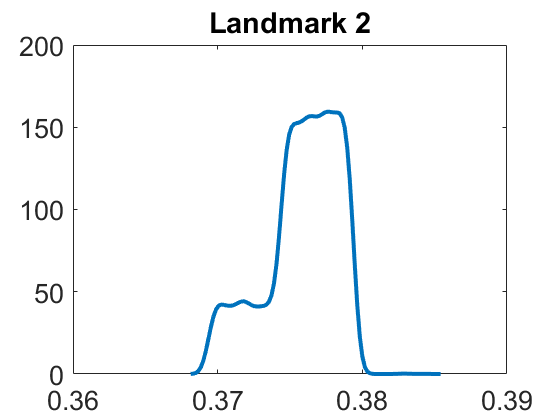}&\includegraphics[width=0.82in]{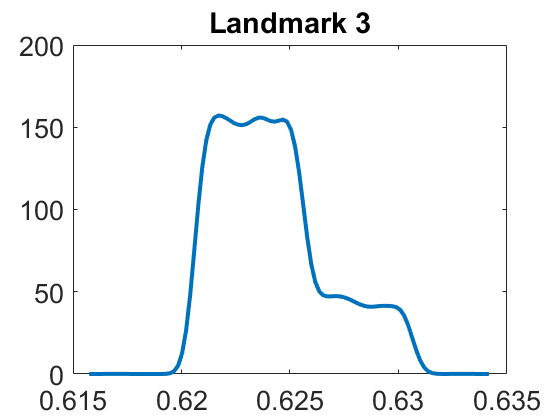}&\includegraphics[width=0.82in]{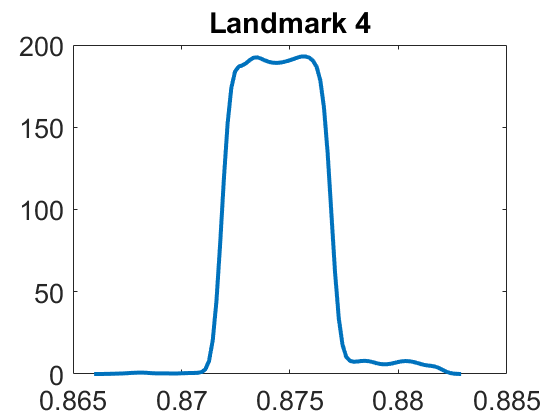}\\
    \hline
    \end{tabular}
\end{center}
\caption{Left: Re-scaled and rotated versions of $\beta$ with posterior landmark locations obtained using MCMC: red = 1, yellow = 2, purple = 3, green = 4. Right: Density plots of marginal posterior samples $\theta_i|\beta^{(N)},\ i=1,\ldots,4$.}
\label{fig:Experiment}
\end{figure}

It is important to check that our inference in Section \ref{OSC} is invariant to shape-preserving transformations, which include translation, scaling and rotation; re-parameterization is not considered here, as the given curves are always sampled using arc-length (due to the population homogeneity assumption). Our models are automatically invariant to translations due to the model's dependence on the SRVF only, which is translation invariant as it is defined using the derivative of $\beta$. A re-scaling of the curve should also result in no change to inference, as curves are pre-processed to have unit length. Figure \ref{fig:Experiment} confirms this for $\beta$ scaled by a factor of two; the resulting marginal posteriors look extremely similar to those of the original curve.  The invariance to rotations is not immediately obvious. In Figure \ref{fig:Experiment}, we also demonstrate inference on a version of the original curve $\beta$ which was rotated by 45 degrees counter-clockwise.  The marginal posteriors again appear to coincide with the original densities, and landmarks are located at the peaks and valleys as before.  These experiments were run under the same settings as the original simulated curve example, and demonstrate that the proposed Bayesian model is invariant to all relevant shape-preserving transformations.

\subsection{Sensitivity Analysis to Choice of Hyperparameters}	\label{Sens}
As with any Bayesian analysis, studying the sensitivity of inference to the choice of hyperparameters is an important consideration.  In our model, we feel that it is important to assess the impact of $a$ and $b$ that appear in the prior on the nuisance parameter $\kappa$. These two hyperparameters play prominent roles in the marginalized likelihood of $\boldsymbol{\beta}^{(N)}$.  Our goal is to select priors that are weakly informative.  In particular, since $a$ is absorbed into an exponent in $f(\boldsymbol{\beta}^{(N)}|\boldsymbol{s})$ (which involves the number of points $N$ and the number of curves $M$), reasonably low choices of $a$, as compared to the product of $N$ and $M$, will not result in much change in the analysis.  However, the choice of $b$ is much more impactful, as large values of $b$ will tend to dominate the linear reconstruction error term; this will result in a flattening of the posterior, and greater variability in inference of $\boldsymbol{\theta}|\boldsymbol{\beta}^{(N)}$.

Table \ref{tab:Sensitivity} shows marginal 95\% posterior credible intervals of the components of $\boldsymbol{\theta}$ for the example described in Section \ref{OSC} under different prior hyperparameter settings (as compared to the original setting of $a=1,\ b=0.01$).  As expected, the intervals remain very similar when $a$ is changed.  In a similar fashion, as $b$ is decreased toward 0 (and the prior provides less influence on posterior inference), the credible intervals change very little from the ones obtained under the original setting.  However, as $b$ is increased, we are incorporating more information into the prior: the magnitude of $b$ becomes more comparable to that of the linear reconstruction error. This increases the variance in the posterior and results in wider credible intervals for landmark locations. Nonetheless, as is evident in this table, our overall inference is robust to the choice of hyperparameters $a$ and $b$.

\begin{table}
\begin{center}
\begin{tabular}{|c|c|c|c|c|c|}
\hline
$a$ & $b$ & $\theta_1$ & $\theta_2$ & $\theta_3$ & $\theta_4$\\
\hline
1 & 0.01 & $(0.1215,0.1280)$ & $(0.3699,0.3792)$ & $(0.6208,0.6297)$ & $(0.8720,0.8781)$\\
\hline
0.01 & 0.01 & $(0.1217,0.1280)$ & $(0.3700,0.3793)$ & $(0.6208,0.6300)$ & $(0.8720,0.8782)$\\
0.1 & 0.01 & $(0.1219,0.1280)$ & $(0.3699,0.3793)$ & $(0.6208,0.6301)$ & $(0.8720,0.8780)$\\
3 & 0.01 & $(0.1227,0.1280)$ & $(0.3699,0.3792)$ & $(0.6208,0.6300)$ & $(0.8720,0.8782)$\\
5 & 0.01 & $(0.1226,0.1280)$ & $(0.3700,0.3792)$ & $(0.6208,0.6301)$ & $(0.8720,0.8776)$\\
\hline
1 & 0.0001 & $(0.1231,0.1280)$ & $(0.3700,0.3793)$ & $(0.6208,0.6300)$ & $(0.8720,0.8769)$\\
1 & 0.001 & $(0.1230,0.1280)$ & $(0.3700,0.3792)$ & $(0.6208,0.6300)$ & $(0.8720,0.8769)$\\
1 & 0.1 & $(0.1190,0.1302)$ & $(0.3697,0.3793)$ & $(0.6208,0.6303)$ & $(0.8697,0.8810)$\\
1 & 1 & $(0.1124,0.1377)$ & $(0.3629,0.3876)$ & $(0.6123,0.6381)$ & $(0.8621,0.8882)$\\
\hline
\end{tabular}
\end{center}
\caption{95\% credible intervals for $\boldsymbol{\theta}|\beta^{(N)}$ under different choices of prior hyperparameters for $\kappa$, for the simulated curve example in Section \ref{OSC}.}
\label{tab:Sensitivity}
\end{table}

\subsection{Inference Comparison of the Number of Landmarks}	\label{CompInf}
\begin{figure}[!t]
\begin{center}
\begin{tabular}{|c|c|}
\hline
\includegraphics[width=2in]{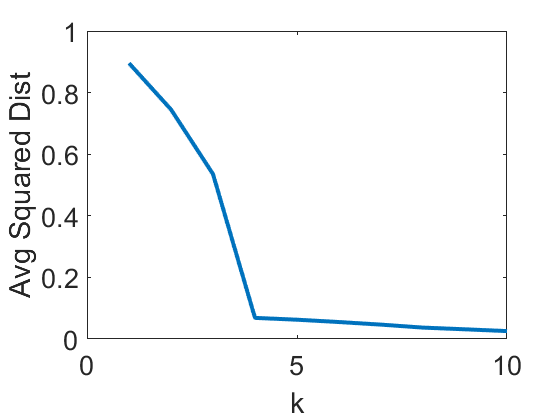}&\includegraphics[width=2in]{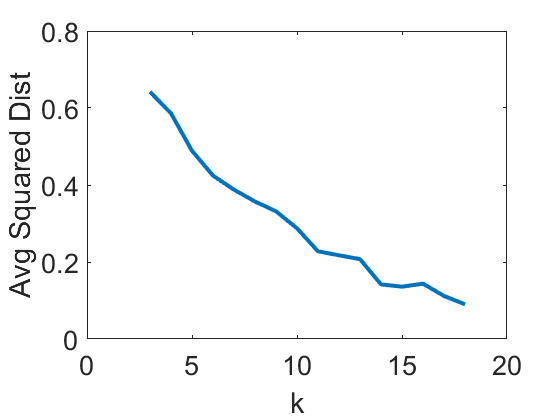}\\
\hline
\end{tabular}
\end{center}
\caption{Average cumulative squared distance vs. $k$ using MCMC sampling for the simulated curve from Section \ref{OSC} (left) and the deer from Section \ref{Complex} (right).}
\label{fig:dsCrit}
\end{figure}

In Section \ref{OSC}, we assumed $k=4$ based on the number of significant features of $\beta$. However, perhaps it is ideal to use fewer or more landmarks based on the reconstruction error. First, we select $k$ using the distance criterion from Section \ref{ASDC}. The left panel of Figure \ref{fig:dsCrit} shows the plot of the average cumulative squared distance ($d_k^2$) as a function of $k$ for $k=1,\ldots,10$; we used 100,000 iterations of MCMC with $\ a=1,\ b=0.01$, and now $N=100$ to generate posterior samples for each value of $k$. As expected, the average cumulative squared distance decreases as $k$ increases due to the reduced reconstruction error. However, at $k=4$, we observe a clear ``elbow" of the curve after which the marginal utility of adding additional landmarks is diminished. Thus, it would appear reasonable to select $k=4$ based on this criterion. However, as stated earlier, this process requires the user to identify this point on the curve, which may not always be obvious, and removes the automation in landmark detection.

\begin{figure}[!t]
\begin{center}
\begin{tabular}{|c|c|c|c|}
\hline
$\lambda=10^{-6}$ & $\lambda=10^{-5}$ & $\lambda=0.1$ & $\lambda=1$\\
\hline
\includegraphics[width=1.4in]{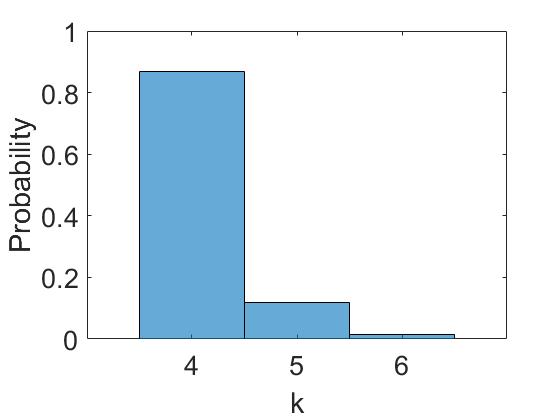}&\includegraphics[width=1.4in]{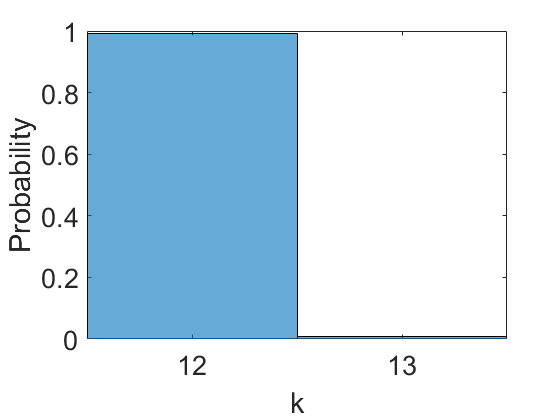}&\includegraphics[width=1.4in]{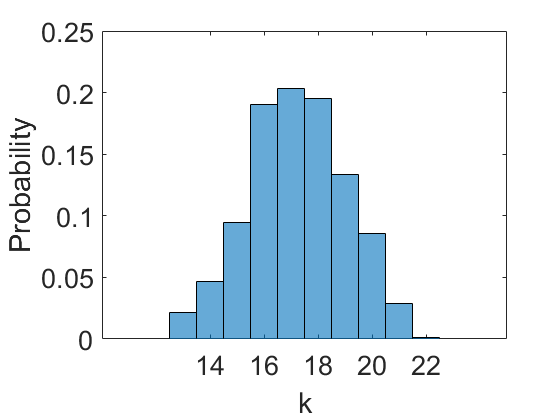}&\includegraphics[width=1.4in]{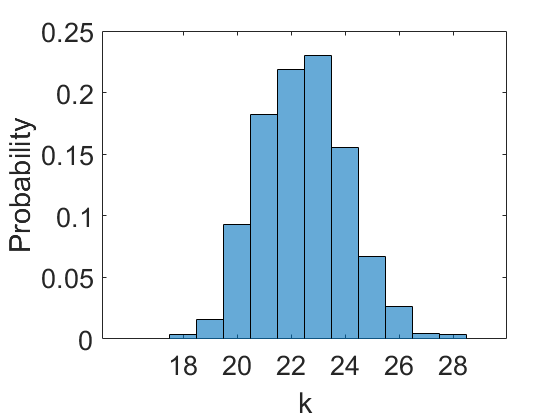}\\
\hline
\includegraphics[width=1.4in]{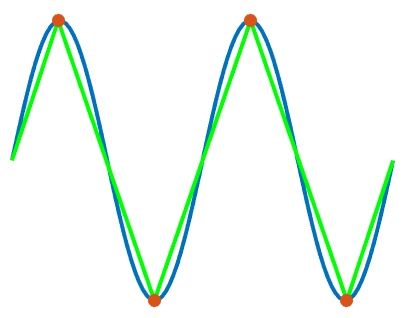}&\includegraphics[width=1.4in]{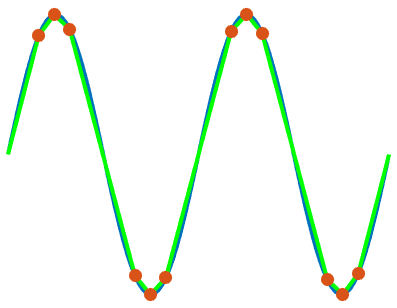}&\includegraphics[width=1.4in]{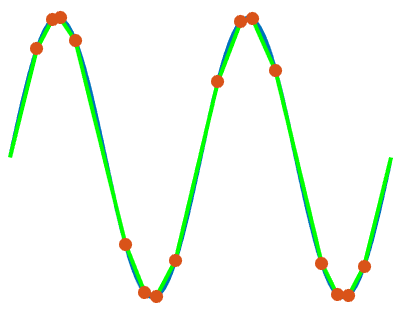}&\includegraphics[width=1.4in]{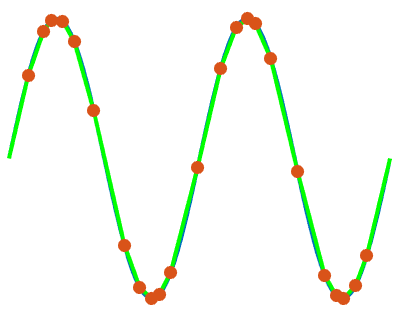}\\
\hline
\end{tabular}
\end{center}
\caption{Inference on the number of landmarks and their locations for the simulated curve example from Section \ref{OSC}; Top: Histograms of samples of $k|\beta^{(N)}$ for different settings of $\lambda$, obtained using RJMCMC. Bottom: Conditional on the mode of the posterior of $k$, linear reconstructions (green) of $\beta$ (blue) based on the mean configuration (red) under different values of $\lambda$.}
\label{fig:kToy}
\end{figure}

In comparison to the user-selected distance criterion, now we let $k$ be inferred using RJMCMC with the conditional model, as outlined in Section \ref{RJMCMC}. We use Algorithm 1 (described in Supplementary Materials) with the concentration parameter $\alpha=1$ for the prior on $\boldsymbol{s}$. We select $v=0.02$ for the variance of the normal proposal in the ``Stay" step, and set $a=1$, $b=0.01$ for the prior on $\kappa$.  Again, $N=100$ evaluation points are used.  As stated in Section \ref{ext}, varying $\lambda$ (the prior parameter for $k$) changes the magnitude of the penalty on $k$. Thus, we present a path of posterior solutions for different values of $\lambda$, as listed in Figure \ref{fig:kToy}.  After running the algorithm for 100,000 iterations, we discard the first 10,000 iterations as burn-in, and take every 100th iteration to reduce autocorrelation and form the approximate posterior distribution. Convergence is diagnosed by monitoring acceptance rates, the log posterior, and examining trace plots of the parameters given values of $k$. The top panel of Figure \ref{fig:kToy} shows posterior histograms for $k|\beta^{(N)}$ for the different settings of $\lambda$.  

As expected, as $\lambda$ increases, the penalty for large values of $k$ diminishes, and thus the posterior of $k|\beta^{(N)}$ is shifted toward higher values.  Note that $\lambda=10^{-6}$ yields a posterior mode which is consistent with the $k$ obtained using the criterion-based approach. Posteriors with larger values of $k$ tend to exhibit greater variability as well, since even miniscule differences in linear reconstruction error are rewarded when $\lambda$ is not extremely small.  Thus, controlling $\lambda$ allows the user to select how detailed these linear reconstructions need to be to represent the given data: large $\lambda$ will favor reconstructions which capture the majority of the high curvature points (i.e., small-scale details), while small $\lambda$ aims for reconstructions which are more parsimonious and ignore smaller details of the shape.  The magnitude of $\lambda$ is dependent on the number of evaluation points $N$ and curves $M$; more of either requires a ``stricter" regularization (i.e., $\lambda$ must be made much smaller to have a penalizing impact on the likelihood).  This is shown in the bottom panel of Figure \ref{fig:kToy}. Here, linear reconstructions of $\beta$ are shown for values of $k$ which exhibit high posterior probability under various settings of $\lambda$.  The displayed reconstructions are formed from the posterior mean of $\boldsymbol{\theta}|\beta^{(N)}$ for the particular value of $k$.  Notice that additional landmarks are placed around the detailed peaks and valleys, which are much more crucial to the linear reconstruction than the other parts of $\beta$.

\subsection{Inference Based on Multiple Curves}	\label{CompMult}
\begin{figure}[!t]
\begin{center}
\begin{tabular}{|c|}
\hline
\includegraphics[width=2.1in]{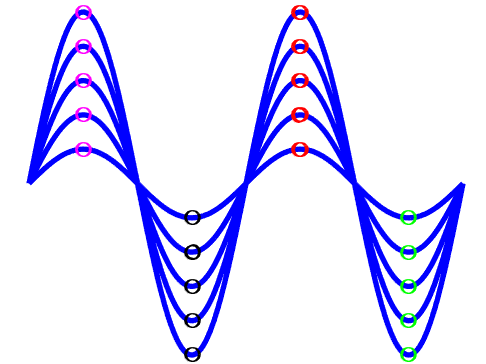}\\
\hline
\end{tabular}
    \begin{tabular}{|c|c|}
    \hline
    \includegraphics[width=1in]{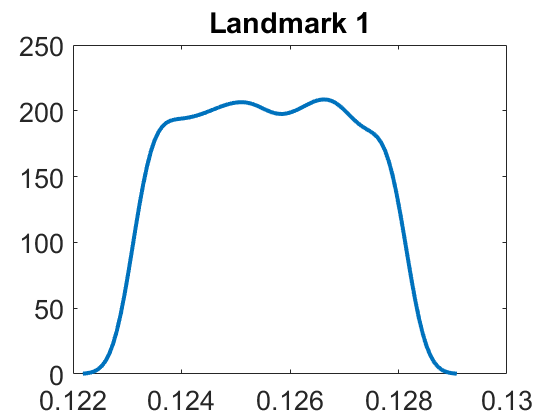} & \includegraphics[width=1in]{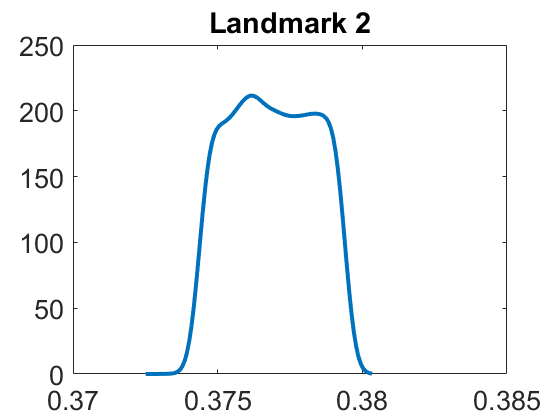}\\
    \hline
    \includegraphics[width=1in]{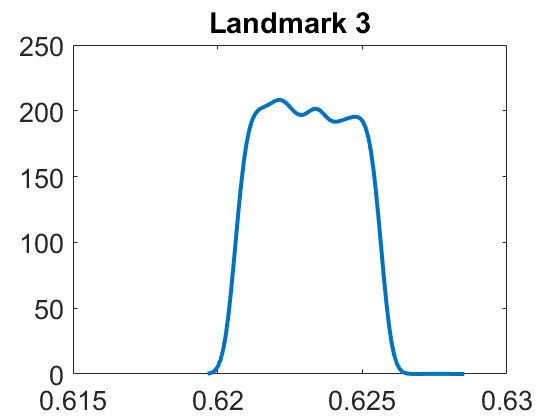} & \includegraphics[width=1in]{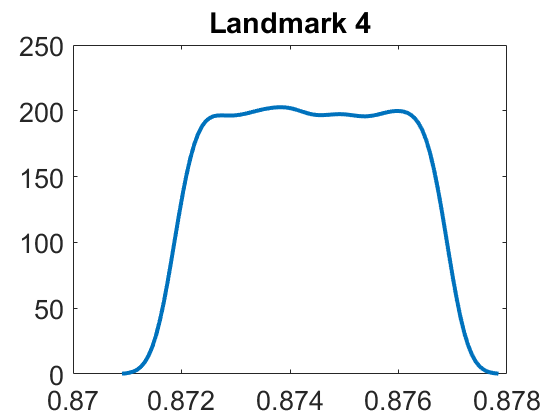}\\
    \hline
    \end{tabular}
\end{center}
\caption{Left: Curves $\beta_1,\ldots,\beta_5$ with posterior landmark locations obtained from MCMC: magenta = 1, black = 2, red = 3, green = 4. Right: Density plots of posterior samples $\theta_i|\boldsymbol{\beta}^{(N)}, \ i=1,\ldots,4$.}
\label{fig:5ToyExLand}
\end{figure}

Most problems of landmark detection involve multiple curves; the proposed model is also applicable to this scenario.  Consider a collection of $M=5$ curves (with $N=200$ evaluation points), each of which has two peaks and two valleys, but now with different heights: $\beta_m(t) = [t,\ m \sin(4\pi t)]^\top,\ 0 \leq t \leq 1,\ m=1,\ldots,5$. Our goal is to infer the locations of $k=4$ landmarks, as in Section \ref{OSC}. Because the peaks and valleys occur at the same locations along each curve, we expect improved inference of landmark locations through more precise estimation of landmarks (as compared to that in Section \ref{OSC}), since their number and locations are common between curves. However, this is not immediately obvious, because while the number of curves increased from the previous example, the cumulative linear reconstruction error will also increase. The proposal $h$ (from Section \ref{MCMC}) is chosen to be a normal density centered at the previous value of the chosen component, with variance $v=0.02$. We obtain $10^6$ dependent samples via MCMC, and the approximate posterior is again formed by discarding the first $100,000$ iterations for burn-in and thinning by every $100$ iterations. Figure \ref{fig:5ToyExLand} shows the posterior landmark locations plotted on all five curves (left) as well as density estimates for $\boldsymbol{\theta}$ (right). These distributions are slightly more concentrated than those in Figure \ref{fig:ToyExLand} as a result of the increased sample size.

\section{Applications}	\label{Apps}
\subsection{Complex Shapes in Computer Vision}	\label{Complex}
In this section, we present examples of posterior landmark inference applied to complex shapes from the MPEG-7 dataset\footnote{http://www.dabi.temple.edu/~shape/MPEG7/dataset.html}, a well-known dataset of shapes in computer vision. These shapes have been extracted from binary images. All of these examples involve closed curves, so we perform the additional pre- and post-processing steps as described in Section \ref{Imp}.

\begin{figure}
\begin{center}
\begin{tabular}{|c|c|}
\hline
\includegraphics[width=1.5in]{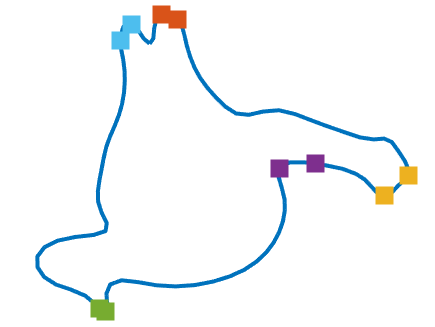}&\includegraphics[width=1.5in]{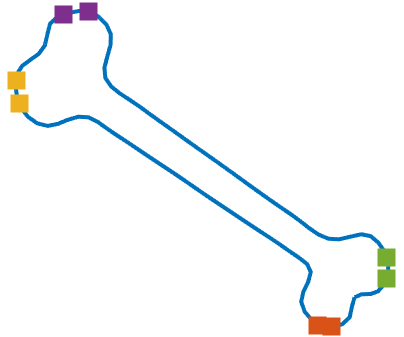}\\
\hline
\end{tabular}
\end{center}
\caption{Posterior 95\% credible intervals for landmarks on the bird (left) and bone (right). Colors match for each component of $\boldsymbol{\theta}$: red = 1, yellow = 2, purple = 3, green = 4, blue = 5.}
\label{fig:Bird}
\end{figure}

The left panel of Figure \ref{fig:Bird} shows the outline of a bird, which contains multiple detailed parts. The area around the feet complicates linear reconstructions using small landmark sets. We select $k=5$ with the same MCMC settings as for the simulated curve example in Section \ref{OSC}, except with proposal variance $v=0.04$. The 95\% credible intervals are shown on the bird's outline. These intervals are very narrow in general, and appear to capture the extreme points of the outline which help minimize the reconstruction error. Notice that the beak, which is an important feature but quite isolated from the other prominent features, has a very narrow credible interval. Clearly, this is an important structure that must be captured by the linear reconstruction. The right panel of Figure \ref{fig:Bird} shows the credible intervals for a bone shape with $k=4$. The credible intervals are again quite narrow, indicating that the four extreme points of the bone will yield a good linear reconstruction of the object. In this example, placing additional landmarks at the high absolute curvature points on the bone may be beneficial.

\begin{figure}[!t]
\begin{center}
\begin{tabular}{|c|c|}
\hline
\includegraphics[width=2.2in]{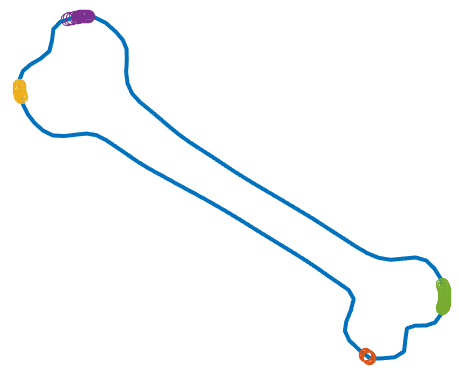}&\includegraphics[width=2.5in]{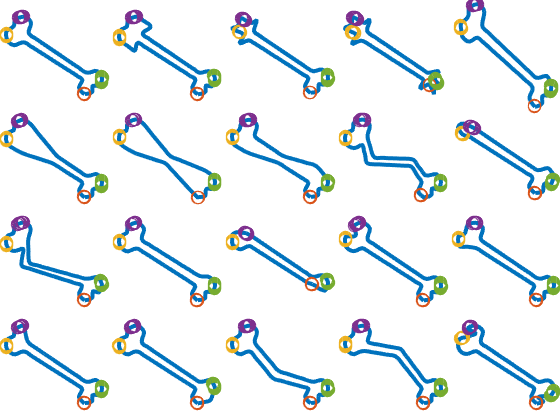}\\
\hline
\end{tabular}
\end{center}
\caption{Posterior landmark locations plotted on the extrinsic sample mean of the 20 bones (left) and each individual bone (right); red = 1, yellow = 2, purple = 3, green = 4.}
\label{fig:20Bones}
\end{figure}

The MPEG-7 dataset features $M=20$ observations of each type of shape. To examine joint inference for multiple shapes, we take a further look at posterior samples drawn for $M=20$ bones with $k=4$ landmarks. Figure \ref{fig:20Bones} shows posterior landmark locations on the extrinsic mean (found simply by averaging all coordinate pairs at each of the $N$ evaluation points) of the 20 bones, as well as on each individual bone. Notice that the model still captures landmarks at the high absolute curvature points of the bone, even when there are abnormalities within an individual bone structure.

\begin{figure}[!t]
\begin{center}
\begin{tabular}{|c|cc|}
\hline
$M=1$ & \multicolumn{2}{|c|}{$M=2$}\\
\hline
\includegraphics[width=1.4in]{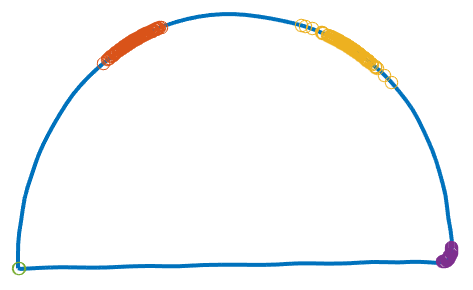}&\includegraphics[width=1.5in]{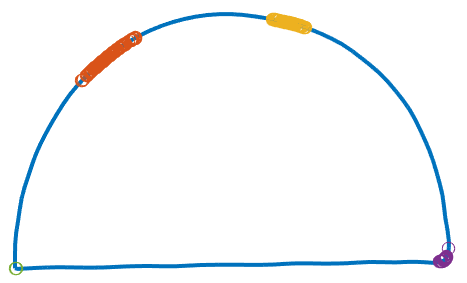}&\includegraphics[width=1.4in]{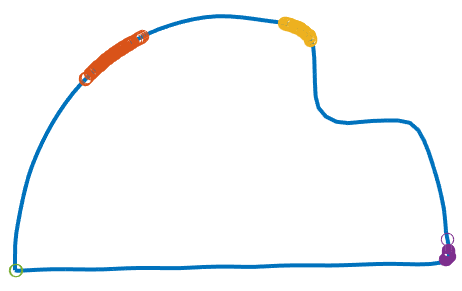}\\
\hline
\end{tabular}
\end{center}
\caption{Posterior inference for $k=4$ landmarks for one half circle (left) and two half circles (right). Notice the change in inference for landmark 2; red = 1, yellow = 2, purple = 3, green = 4.}
\label{fig:HalfCirc}
\end{figure}

An interesting example to consider is the introduction of a second shape to joint posterior inference, where the second shape has a fairly different structure. Consider Figure \ref{fig:HalfCirc}; on the left are posterior locations of $k=4$ landmarks on a half circle (where $M=1$).  Notice the low variability in the landmarks on the base, and higher variability in landmarks on the curved portion of the shape, reflecting the difficulty of a full linear reconstruction which captures the shape's curvature.  If a second half circle is introduced (with a large portion of the right side ``missing"), and posterior sampling is done for $M=2$ (as shown in the right panel of the figure), then inference of the landmarks on the top portion of the shapes changes drastically. In particular, landmark 2 (in yellow) shifts locations slightly as compared to in the $M=1$ case, and exhibits much lower variability. This is due to the large amount of curvature that occurs in the newly introduced shape, which forces a linear reconstruction to capture that particular feature.

\begin{figure}[!t]
\begin{center}
\begin{tabular}{|c|c|c|}
\hline
$\lambda=0.00001$ & $\lambda=0.0001$ & $\lambda=0.001$\\
\hline
\includegraphics[width=1.5in]{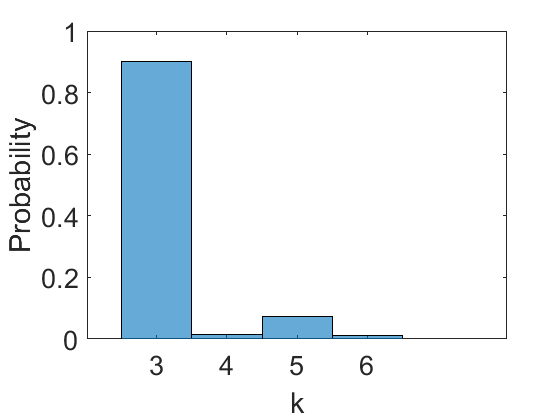}&\includegraphics[width=1.5in]{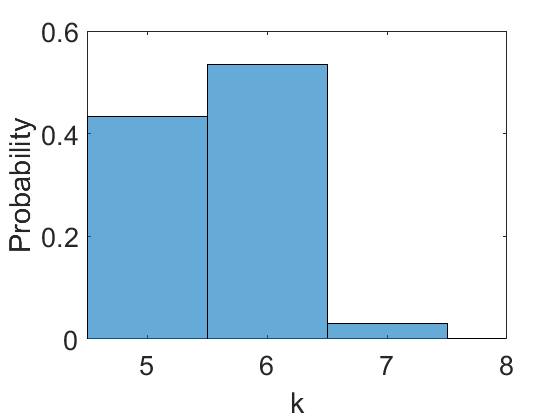}&\includegraphics[width=1.5in]{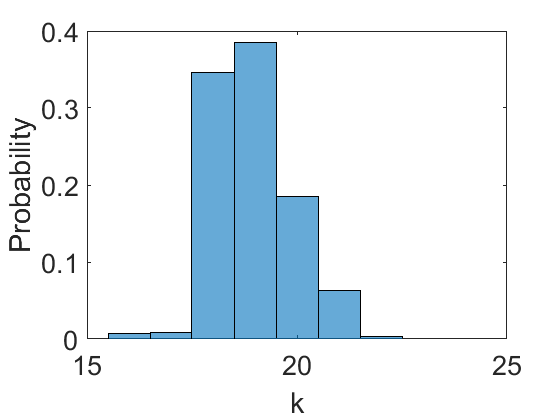}\\
\hline
\includegraphics[width=1.5in]{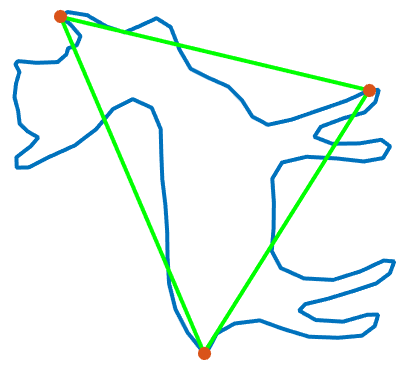}&\includegraphics[width=1.5in]{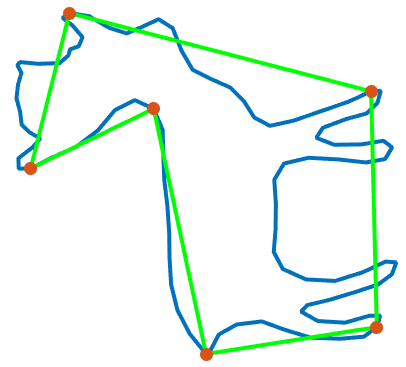}&\includegraphics[width=1.5in]{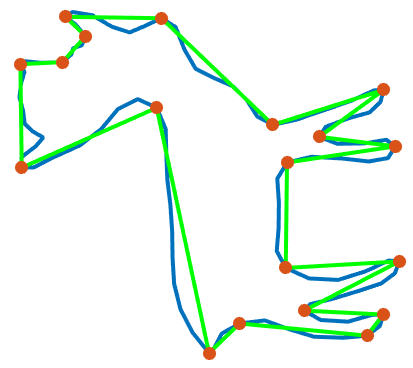}\\
\hline
\end{tabular}
\end{center}
\caption{Top: Histograms of samples of $k|\beta^{(N)}$ for different settings of $\lambda$, obtained using RJMCMC. Bottom: Conditional on the mode of the posterior of $k$, linear reconstructions of $\beta$ based on the median configuration under different values of $\lambda$.}
\label{fig:kDeer}
\end{figure}

For shapes which are even more complex, selecting the number of landmarks is not trivial; consider the deer in Figure \ref{fig:kDeer}. Between the legs and antlers, the researcher will find it difficult to select an appropriate number of landmarks heuristically; thus, it makes sense to assume $k$ as unknown. We first attempt to estimate $k$ for the deer outline using the distance criterion from Section \ref{ASDC}; the result is shown in the right panel of Figure \ref{fig:dsCrit}.  While in the simulated open curve example there was a discernible point for which there was no benefit to adding more landmarks, there does not seem to be such a clear distinction in this case, even as we increase the number of landmarks to more than $15$. As mentioned previously, this can happen, particularly with complex objects, because it is not as clear where or how many landmarks should be selected. This plot will begin to level off, as increasing the number of landmarks will certainly improve the linear reconstruction; it is possible that the ``elbow" point has simply not occurred yet when $k=18$.  Computationally, this is extremely inefficient, and thus, it may make more sense to proceed with this problem by estimating $k$ within the Bayesian model.

Next, we use RJMCMC to estimate the number of landmarks $k$ on the deer example with $\alpha=1,\ v=0.05,\ a=1$, and $b=0.01$. As in the simulated open curve example in Section \ref{CompInf}, Figure \ref{fig:kDeer} shows posterior summaries of $k|\beta^{(N)}$ for different values of $\lambda$, which again acts as a regularizer. As expected, increasing $\lambda$ shifts the marginal posterior of $k$ to higher values, and rewards better reconstructions over sufficiently small values of $k$. Linear reconstructions for the median configuration of landmarks are also shown in Figure \ref{fig:kDeer}. Notice that all three of the landmark configurations capture important features of the deer outline. In fact, our approach allows the user to control the number of landmarks selected on the shape of interest through an appropriate choice of $\lambda$. For complex shapes, such as the deer example given here, it may be beneficial to select more landmarks; on the other hand, for simpler shapes like the simulated example, a few landmarks are sufficient.

\subsection{Mice Vertebrae}	\label{mice}
Biology is a useful application of automatic landmark detection, as existing approaches usually rely on expert knowledge. The second thoracic mice vertebrae exhibits differences in shape and size when mice are controlled for diet. In this section, we use data obtained from the R `shapes' package, as described in \cite{DrydenBook}. Refer to Figure 3 of \cite{LCESA} for a description of the anatomy of the mouse vertebra.

\begin{figure}[!t]
\begin{center}
\begin{tabular}{|c|c|}
\hline
& \includegraphics[width=1.8in]{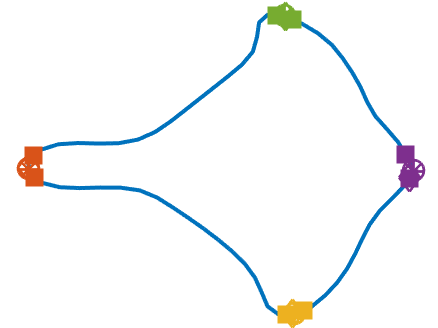}\\
\hline
$\overline{\boldsymbol{\theta}}$ & $(0.2935,0.5999,0.7843,0.9836)$\\
\hline
$\textrm{med}(\boldsymbol{\theta})$ & $(0.2947,0.5997,0.7839,0.9835)$\\
\hline
$\boldsymbol{\theta}_{MAP}$ & $(0.2927,0.5976,0.7760,0.9822)$\\
\hline
\end{tabular}
\begin{tabular}{|cc|}
\hline
\includegraphics[width=1.2in]{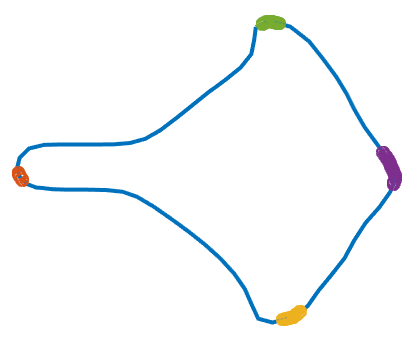}&\includegraphics[width=1.2in]{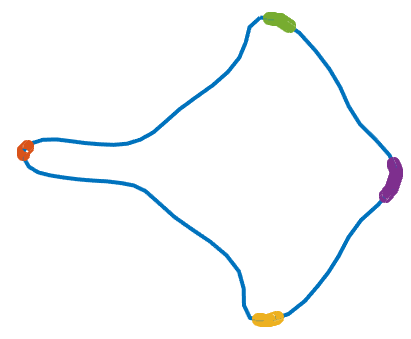}\\
\includegraphics[width=1.2in]{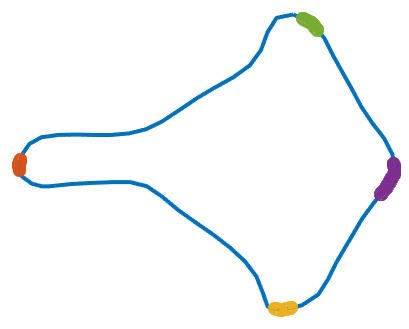}&\includegraphics[width=1.2in]{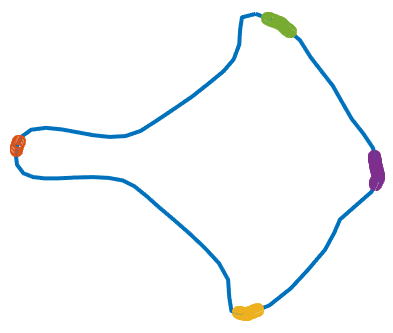}\\
\hline
\end{tabular}
\end{center}
\caption{Left: Mouse vertebra, with the posterior mean, median, and 95\% credible intervals of $\boldsymbol{\theta}|\beta^{(N)}$: circle = mean, asterisk = median, diamond = MAP, squares = 95\% credible interval. Right: Posterior sample landmark locations for $M=4$ mice vertebrae. The colors match for each component of $\boldsymbol{\theta}$: red = 1, yellow = 2, purple = 3, green = 4.}
\label{fig:MicePost}
\end{figure}

We first begin by analyzing a single mouse vertebra; the outline appears to have four distinct landmark locations (which correspond to the neural spine, centrum, and transverse processes), so we assume $k=4$ and proceed with the fixed $k$ model.  Once again, we perform random walk Metropolis, using the same settings as in Section \ref{OSC}.  The resulting posterior summaries are shown in Figure \ref{fig:MicePost}.  Notice that the 95\% credible interval is quite narrow for estimating all of the landmarks, which appear to correspond to the neural spine, centrum, and transverse processes, all of which have anatomical meaning.

To see the impact of increased sample size on posterior inference, we also show results for $M=4$ closed curves of mice vertebrae outlines. We use $10^6$ iterations of MCMC with appropriate burn-in and thinning, and the same model parameters as in the $M=1$ case presented earlier (with $N=61$). The posterior landmark sample locations are plotted on the right side of Figure \ref{fig:MicePost}. Notice that, similarly to the $M=5$ sine curve example of Section \ref{CompMult}, the variability in landmark locations is somewhat smaller when inference is based on $M=4$ mice vertebrae rather than one; this is due to the likelihood being much more concentrated due to its dependence on the sum of interpolated distances over the curves in the data. The estimated landmark locations appear to identify the four natural landmarks of the vertebrae (neural spine, transverse processes and centrum).

\begin{figure}[!t]
\begin{center}
\begin{tabular}{|c|}
\hline
\includegraphics[width=1.9in]{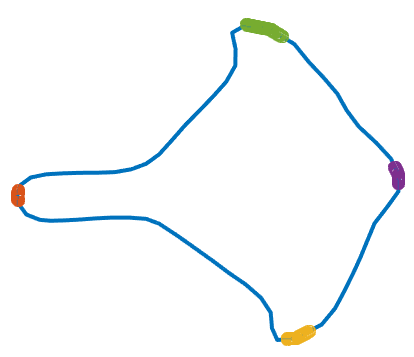}\\
\hline
\end{tabular}
\begin{tabular}{|c|c|}
\hline
    \includegraphics[width=1.1in]{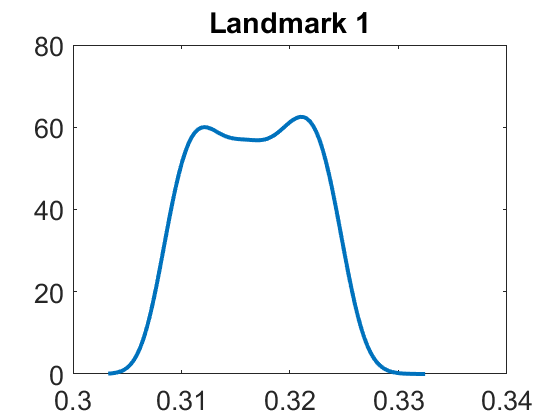} & \includegraphics[width=1.1in]{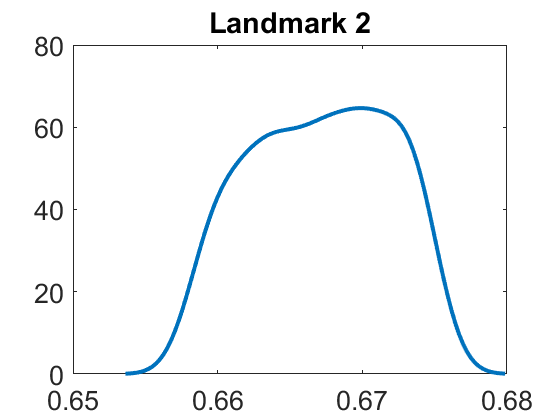}\\
    \hline
    \includegraphics[width=1.1in]{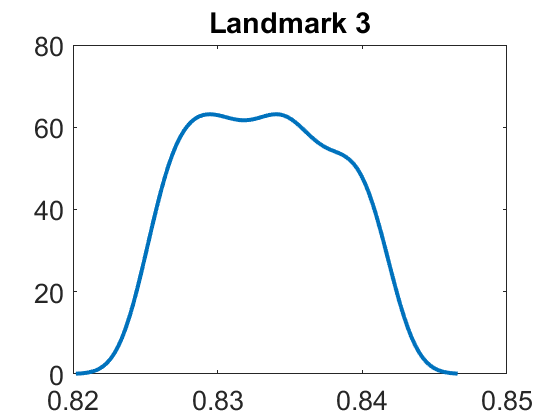} & \includegraphics[width=1.1in]{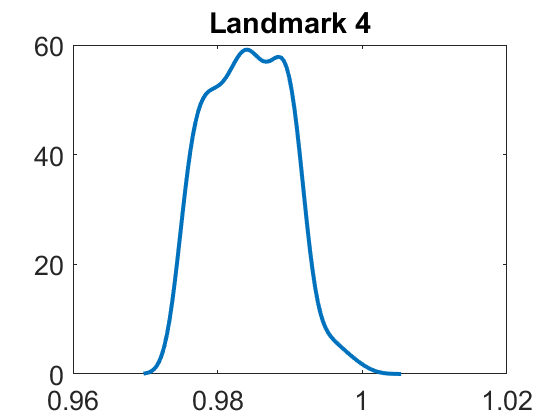}\\
\hline
\end{tabular}
\end{center}
\caption{Left: Extrinsic mean curve $\bar{\boldsymbol{\beta}}$ for 30 control mice with posterior landmark locations obtained from MCMC: red = 1, yellow = 2, purple = 3, green = 4.  Right: Density plots of marginal posterior samples $\theta_i|\boldsymbol{\beta}^{(N)}, \ i=1,\ldots,4$.}
\label{fig:LotsofMice}
\end{figure}

This particular dataset features mice controlled for diet. We now examine posterior landmark locations for the full sample size ($M=30$) of vertebrae from a subpopulation of mice which were not genetically selected for a large or small body weight (i.e., a control group of mice). The same MCMC settings as above are used (except with $10^5$ iterations), and results are shown in Figure \ref{fig:LotsofMice}.  Due to the large number of samples, we show a plot of landmark locations on the extrinsic mean $\bar{\boldsymbol{\beta}}$ of the sample of 30 curves $\{\beta_1,\ldots,\beta_{30}\}$ (found simply by averaging all coordinate pairs at each of the $N$ evaluation points).  As expected, landmarks are again identified at the same locations.  Density plots of posterior samples are also shown in the figure.

\subsection{Brain Substructures from Magnetic Resonance Image Slice} \label{Brain}
As mentioned in Section \ref{Intro}, a particular motivation for automated landmark detection arises in the field of medical imaging. Doctors are often required to manually annotate images of anatomical structures with important landmarks, which is tedious and prone to human error. In this section, we apply our model to four different substructures (caudate, hippocampus, putamen, and thalamus) extracted from brain magnetic resonance images (MRI) of ten different subjects. An example of a subject's original MRI slice, as well as outlines of substructures associated with all ten subjects can be found in \cite{SPIE}.

\begin{figure}
\begin{center}
\begin{tabular}{|c|c|c|}
\hline
\begin{sideways}Caudate \end{sideways}&\includegraphics[width=1.3in]{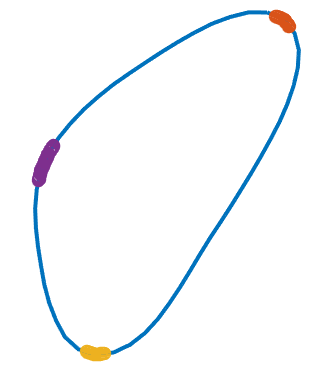}&\includegraphics[width=2.2in]{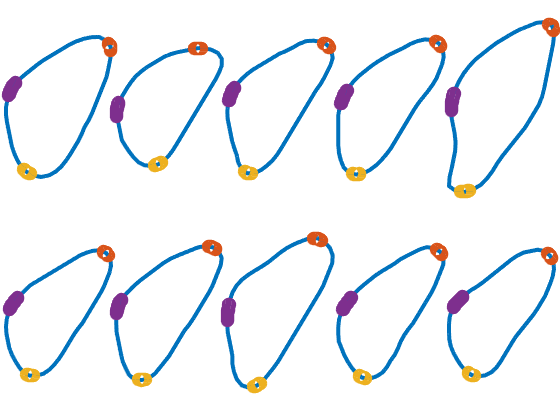}\\
\hline
\begin{sideways}Hippocampus \end{sideways}&\includegraphics[width=1.6in]{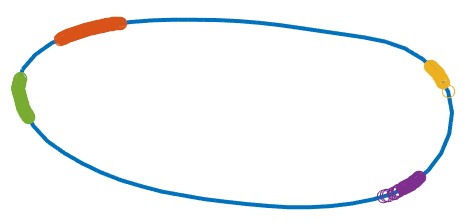}&\includegraphics[width=2.4in]{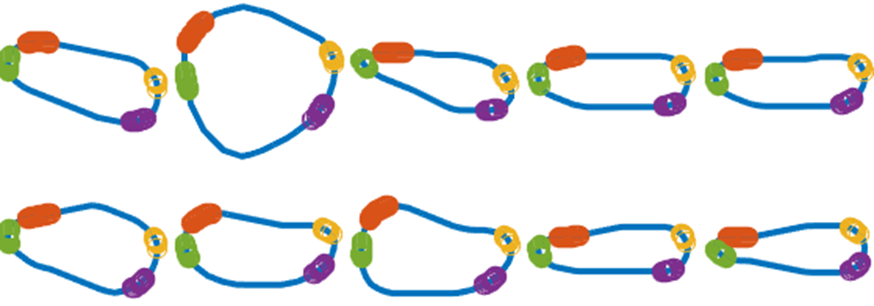}\\
\hline
\begin{sideways}Putamen \end{sideways}&\includegraphics[width=0.8in]{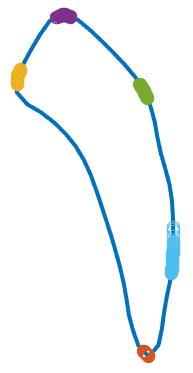}&\includegraphics[width=2.2in]{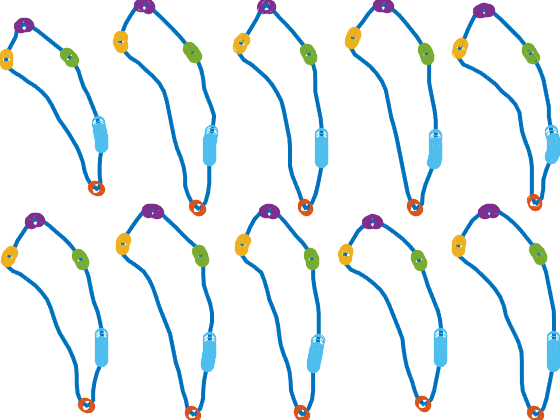}\\
\hline
\begin{sideways}Thalamus \end{sideways}&\includegraphics[width=1.1in]{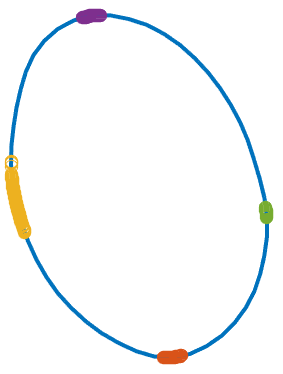}&\includegraphics[width=2.2in]{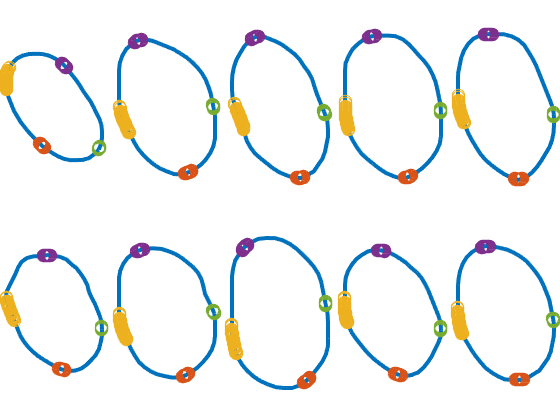}\\
\hline
\end{tabular}
\end{center}
\caption{Posterior landmark locations plotted on the extrinsic sample mean of each substructure (left) and each substructure individually (right) from the sample of ten subjects; red = 1, yellow = 2, purple = 3, green = 4, blue = 5.}
\label{fig:BrainSub}
\end{figure}

Figure \ref{fig:BrainSub} shows posterior landmark locations for the four substructures applied across the sample of $M=10$ subjects; each observation was sampled to $N=50$ points. We choose to demonstrate inference for fixed $k$, where $k=3$ for the caudate, $k=4$ for the hippocampus and thalmus, and $k=5$ for the more structurally-complex putamen. We ran MCMC for $10^6$ iterations for each substructure, using a proposal variance of $v=0.02$; the first $100,000$ iterations were discarded for burn-in, and the remaining sample was thinned by every $100$ iterations. Note the fairly similar amount of variability for all landmarks in both the caudate and hippocampus.  The putamen is interesting, as three landmarks are placed at the top of the structure with low variability, as well as the red landmark located at the bottom of the structure.  However, the fifth landmark (in blue) exhibits more variability, as it does not appear to be as necessary to the linear reconstruction as the other four landmarks. This procedure allows for automatic annotation of landmarks that doctors may otherwise be forced to do manually.

\section{Summary and Future Work} \label{Conc}

We defined a Bayesian model for inference of landmark locations given a set of shapes from a particular population. The benefits of this model include the ability to obtain automatic estimates of landmark locations along with measures of uncertainty, thereby eliminating the need for a researcher to manually annotate important features on shapes. We propose a hierarchical model for both the fixed-landmark and variable-landmark settings, and describe methods for approximate sampling from the posterior distributions. In the variable dimension landmark setting, we discuss the impact of the regularization parameter $\lambda$ on posterior inference on the number of landmarks $k$ for both open and closed curve examples.

One direction for future work is to allow for the assumption of heterogeneous shape subpopulations into the model; at present, it is assumed that all shapes come from a homogeneous population. In the multiple curve case, introducing heterogeneity complicates inference, as values of $\theta$ may not necessarily correspond to the same feature across the sample of shapes (especially in the presence of large elastic variability or missing parts). This can be resolved by first finding the optimal groupwise registration prior to landmark inference (in the manner discussed by \cite{SrivESA}). However, incorporating registration into the Bayesian model by conditioning on a registration function, which respects landmark locations, seems more appropriate. \cite{Cheng} discuss a Bayesian method of function and curve registration without landmarks, which could be extended to include different types of landmark constraints.

We will also explore more efficient posterior sampling strategies. Due to the high-dimensionality of the landmark detection problem, combined with the intricate geometric details of the objects under study, our current MCMC implementation based on component-wise proposals can make it challenging to explore multimodal posteriors. Efficient MCMC schemes, designed to more efficiently traverse multimodal posteriors, may be required for more complex shapes. Finally, we will further explore different choices for model specification. The current model is dependent on the number of evaluation points $N$, which can lead to a highly-peaked likelihood with the potential for multiple modes making posterior inference challenging as described above. A likelihood model which is independent of $N$ has the potential to reduce computational issues.

\newpage

\bibliographystyle{Chicago}
\bibliography{BibALD}

\end{document}